\documentclass{IEEEoj}
\usepackage{cite}
\usepackage{amsmath,amssymb,amsfonts}
\usepackage{algorithmic}
\usepackage{graphicx,color}
\usepackage{textcomp}

\usepackage{url}
\usepackage{hyperref}

\usepackage{booktabs}
\usepackage{subfigure}

\def\BibTeX{{\rmB\kern-.05em \kern-.05em {\sc i\kern-.025em b}\kern-.08em
    T\kern-.1667em\lower.7ex\hbox{E}\kern-.125emX}}
\AtBeginDocument{\definecolor{ojcolor}{cmyk}{0.93,0.59,0.15,0.02}}
\def\OJlogo{\vspace{-14pt}\includegraphics[height=28pt]{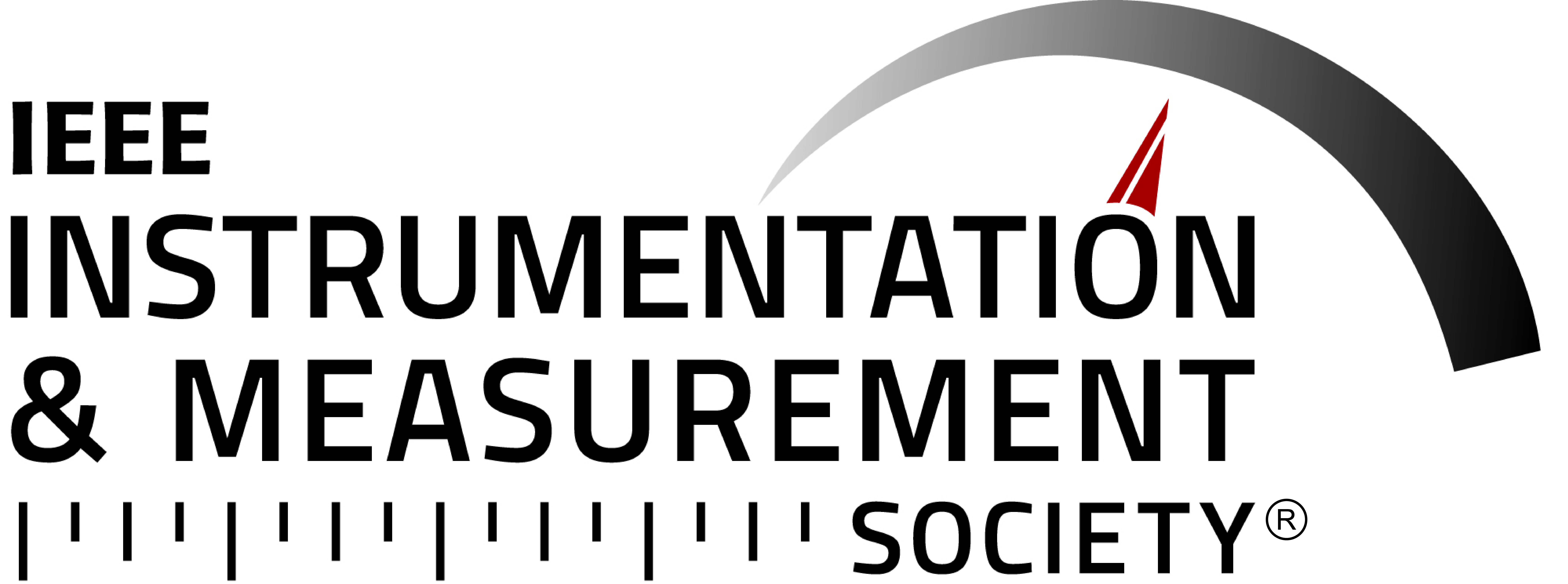}}

\begin{document}

\receiveddate{XX Month, XXXX}
\reviseddate{XX Month, XXXX}
\accepteddate{XX Month, XXXX}
\publisheddate{XX Month, XXXX}
\currentdate{XX Month, XXXX}
\doiinfo{OJIM.2022.1234567}

\title{Financial Crimes in Web3-empowered Metaverse: Taxonomy, Countermeasures, and Opportunities
}
\author{Jiajing Wu\authorrefmark{1}, Senior Member, IEEE, Kaixin Lin\authorrefmark{1}, Dan Lin\authorrefmark{2}, Ziye Zheng\authorrefmark{1}\authorrefmark{3}, Huawei Huang\authorrefmark{2}, Senior Member, IEEE, AND Zibin Zheng\authorrefmark{2}, Fellow, IEEE}
\affil{\authorrefmark{1}School of Computer Science and Engineering, Sun Yat-Sen University, Guangzhou, China.}
\affil{\authorrefmark{2}School of Software Engineering, Sun Yat-Sen University, Guangzhou, China.}
\affil{\authorrefmark{3}School of Software Engineering, South China Normal University, Foshan, China.}

\corresp{CORRESPONDING AUTHOR: Zibin Zheng (e-mail: zhzibin@mail.sysu.edu.cn)).}
\authornote{The work described in this paper is supported by the National Natural Science Foundation of China (61973325, 62032025), the Natural Science Foundations of Guangdong Province (2021A1515011661), and the Guangzhou Basic and Applied Basic Research Project (202102020616).}
\markboth{Financial Crimes in Web3-empowered
Metaverse}{Wu \textit{et al.}}

\begin{abstract}
At present, the concept of metaverse has sparked widespread attention from the public to major industries. With the rapid development of blockchain and Web3 technologies, the decentralized metaverse ecology has attracted a large influx of users and capital. 
Due to the lack of industry standards and regulatory rules, the Web3-empowered metaverse ecosystem has witnessed a variety of financial crimes, such as scams, code exploit, wash trading, money laundering, and illegal services and shops.
To this end, it is especially urgent and critical to summarize and classify the financial security threats on the Web3-empowered metaverse in order to maintain the long-term healthy development of its ecology.
In this paper, we first outline the background, foundation, and applications of the Web3 metaverse. Then, we provide a comprehensive overview and taxonomy of the security risks and financial crimes that have emerged since the development of the decentralized metaverse. For each financial crime, we focus on three issues: a) existing definitions, b) relevant cases and analysis, and c) existing academic research on this type of crime. Next, from the perspective of academic research and government policy, we summarize the current anti-crime measurements and technologies in the metaverse. Finally, we discuss the opportunities and challenges in behavioral mining and the potential regulation of financial activities in the metaverse. 
The overview of this paper is expected to help readers better understand the potential security threats in this emerging ecology, and to provide insights and references for financial crime fighting.

\end{abstract}



\maketitle


\section{Introduction}

Metaverse, literally a combination of the prefix ``meta'' (meaning beyond) and the suffix ``verse'' (abbreviation of ``universe''), describes a world of virtuality and reality beyond the real world built by human beings using digital technology.
Under the context of the \emph{metaverse}, people can get a new Internet experience with high realism and deep immersion.

\begin{figure}
    \centering
    \includegraphics[width=\linewidth]{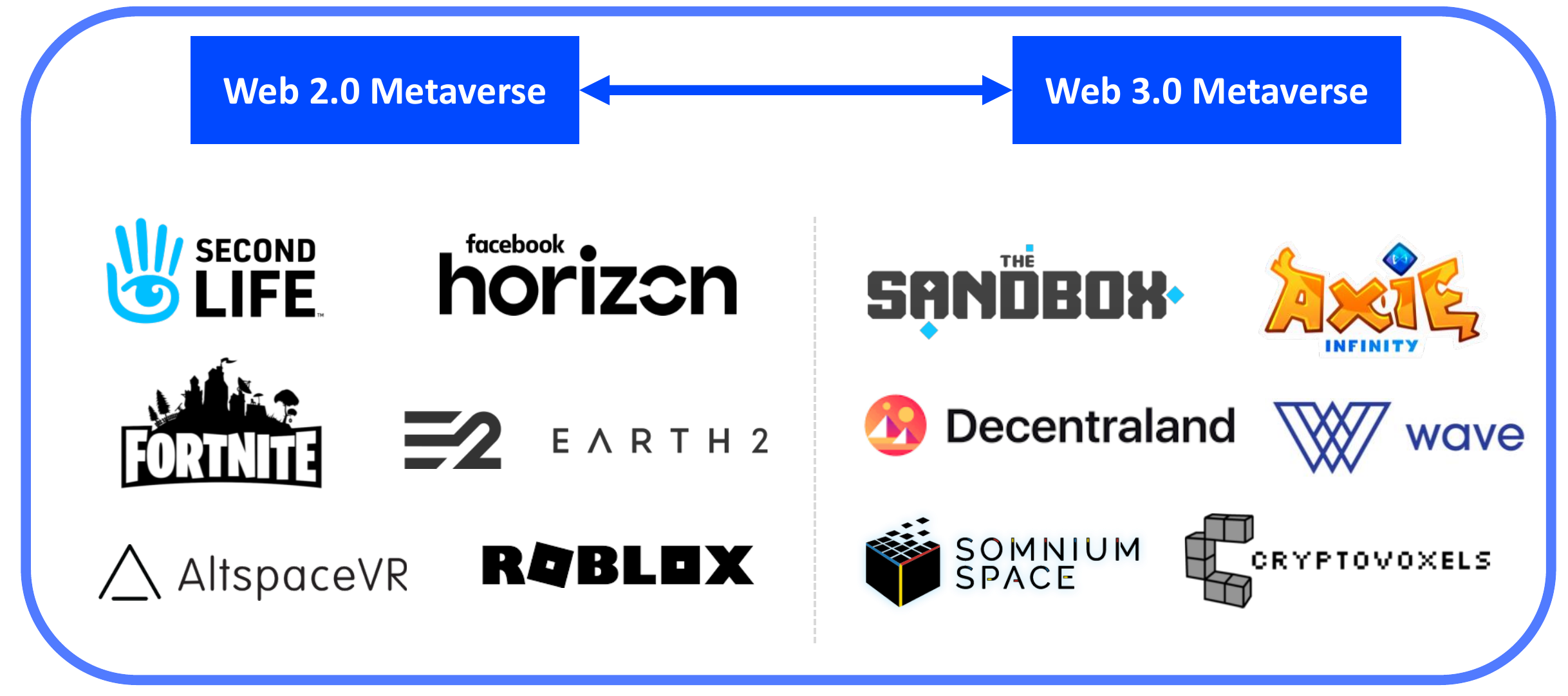}
    \caption{Different types of metaverses.}
    \label{fig:metaverse_type}
\end{figure}

A key consideration in building the metaverse is whether it is centralized (centrally owned and controlled by large technology companies), or decentralized (jointly owned by members of the metaverse community). Typically, the former is referred to as the centralized metaverse, or ``Web2 closed corporate metaverse''; the latter is referred to as the decentralized metaverse, or ``Web3 open crypto metaverse'', as shown in FIGURE~\ref{fig:metaverse_type}.
The concept of ``Web3" here refers to the third iteration of the Internet that has been launched globally in recent years. Compared with the second generation of the Internet, which enables users of the metaverse to ``read and write", Web3 enables users to ``read, write, and own". In the context of ``Web3", users themselves hold the ownership of digital assets and their related derivative powers, which is the technical fundamental for the current decentralized Internet. The decentralized metaverse based on Web3 is referred to as the Web3-empowered metaverse (Web3 metaverse for short), and is the focus of this paper.

In the Web3-empowered metaverse, Web3 and metaverse are mutually supportive and complementary. On the one hand, the metaverse represents the future way of life and business, and provides the upper application scenarios and revolutionary front-end architectures for Web3. 
On the other hand, Web3 is the underlying technology foundation of the decentralized Internet, and provides revolutionary back-end support for the decentralized metaverse ecosystem.
In the metaverse ecosystem, digital creation, digital asset, digital market and digital currency constitute the basic economic system~\cite{yang2022fusing}. If the metaverse is controlled by one company or one organization, the power of its centralized platform may grow wildly, and the economic problems that exist in the centralized real world, such as the uneven distribution of resources and disparity between rich and poor, are likely to be further expanded in the metaverse economic system. Web3, as a new economic infrastructure, shifts trust in complex systems from individual organizations to decentralized nodes and verifiable code. The ``decentralized'' nature of Web3 is believed to help build a more open, autonomous, efficient, and fair metaverse ecosystem.

\textbf{Motivations.}
According to a report of Greyscale~\cite{Grayscale2022Metaverse}, sales of items such as virtual land, goods and services in the Web3 metaverse have exceeded \$200 million.
At the time of writing, the top five metaverse native assets have reached a market value of over \$5 billion\footnote{\url{https://coinmarketcap.com/view/metaverse/}}. 
However, where there is a concentration of value, there is crime, and the Web3 metaverse is no exception. The Web3 metaverse may well be a potentially attractive new vector for financial criminals.
First, the Web3 metaverse will likely inherit and perpetuate the financial crimes that existed in Web3. According to Certik\footnote{\url{https://certik-2.hubspotpagebuilder.com/hack3d-q1-2022-0}}, a blockchain security firm, more than \$2 billion was stolen from Web3 projects as a result of hacking and vulnerabilities in the first half of 2022. Secondly, there are still many uncertainties in the fledgling metaverse, and there are profit--seeking inertia operations such as creating new concepts, speculating on new windfalls, and attracting new investments. Finally, the Web3 metaverse, which lacks industry standards and regulatory rules, may provides a more hidden space for financial crimes, such as fraud, malicious attacks, wash trading, terrorist financing, etc. Therefore, financial regulations needs to be expanded from the real world to the metaverse.

\textbf{Research gaps.} Nowadays, the hotness of the metaverse concept has led to increasing academic interest in the Metaverse as a key research topic. At the time of writing, there have been a number of research and review papers that discuss research issues related to the metaverse from various aspects.
There are some surveys~\cite{sun2022metaverse,wang2022survey,lee2021all,yang2022fusing,ning2021survey,kovacova2022behavioral} that individually or simultaneously present the technical framework of the metaverse, application scenarios, behavioral analysis, security issues, interdisciplinary research opportunities, and the integration of other fields with the metaverse, but these studies do not describe and summarize the current state of the economic system of the metaverse.
Some other reviews pay attention to the economic system of the metaverse and its challenges~\cite{Embracing2022Yuan,belk2022money}, while they mainly focus on the economic logic and business models, and do not discuss the current state of financial crimes or provide anti-crime solutions.
Some recent surveys discuss the crime, scams, and gray economy of the metaverse~\cite{mackenzie2022criminology,smaili2022metaverse}. However, these studies are conducted from the perspective of finance and criminology, and there are no surveys devoted to issues ranging from the technological fundamentals of the metaverse economic system, to financial crimes, and the opportunities and challenges of data-driven metaverse behavior mining and regulation. This motivates us to provide an overview of the concepts, technologies, and applications related to the Web3 metaverse economic system, and to focus on summarizing the possible financial crimes in the metaverse, giving relevant case studies to provide the reader with a comprehensive understanding and key insights.

\textbf{Contributions.} In this paper, we first outline the background, key techniques, and applications related to the the Web3 metaverse, and then provide a comprehensive overview and taxonomy of the security risks and financial crimes that have emerged in the metaverse. Next, we present existing anti-crime related research and measures, and the opportunities and challenges of data-driven metaverse regulation. The contributions of this work are threefold:
\begin{itemize}
    \item We review and summarize the various types of common financial crimes that have emerged since the development of the metaverse in five parts (i.e., scams, code exploits, wash trading, money laundering, and emerging crimes in the metaverse). This provides a reference for regulators, researchers, and practitioners to understand the possible risks of the metaverse economic system.
    \item We investigate the current state of financial crime prevention in the metaverse and its economic system from two perspectives: academic research and policy measurements, to inform investigators and researchers on how to prevent metaverse financial crimes, and to provide strategies and insights for deterring, detecting, and preventing metaverse financial crimes.
    \item We explore the possible opportunities and challenges of data-driven regulation for the metaverse at four levels: source, acquisition, query and indexing, analysis, and applications, and provide guidelines for researchers to design financial behavior mining algorithms and illegal behavior detection techniques for the metaverse.
\end{itemize}

\textbf{Roadmap.} This paper is organized as follows. Section~\ref{sec:RelatedWork} summarizes the current metaverse--related review studies, including the concepts, definitions, applications, security issues and policies related to the metaverse. In Section~\ref{sec:BackgroundandFundamentals}, we describe the relationship between metaverse and Web3, and describe the core framework and technical foundation of Web3 metaverse and typical applications. In Section~\ref{sec:Crimes}, we summarize the possible financial crimes in the Web3 metaverse, giving relevant case studies to provide readers with a comprehensive understanding and key insights. Then, in Section~\ref{sec:Anti-crimes}, we outline the current regulatory policies and countermeasures to deal with financial crimes in the metaverse. Further, we discuss the current data-driven financial regulatory opportunities and and possible challenges in the Web3 metaverse in Section~\ref{sec:OpportunitiesandChallenges}. Finally, we conclude the paper in Section~\ref{sec:Conclusion}. Table~\ref{tab:abbreviations} lists the abbreviations that appear frequently in this paper.

\begin{table}
  \centering
  \caption{List of frequently occurring abbreviations in the alphabetical order.}
    \begin{tabular}{ll}
    \toprule
    \textbf{Acronym} & \textbf{Explanation} \\
    \midrule
    \midrule
    3D    & Three Dimensional \\
    AML   & Anti-money Laundering \\
    AR    & Augmented Reality \\
    DAO   & Decentralized Autonomous Organization \\
    DeFi  & Decentralized Finance  \\
    KYC   & Know Your Customer \\
    NFT   & Non-fungible Token \\
    P2E   & Play-to-earn \\
    VR    & Virtual Reality \\
    Web3 & The Third Iteration of Internet \\
    XR    & Extended Reality \\
    \bottomrule
    \end{tabular}%
  \label{tab:abbreviations}%
\end{table}%

\section{Related Work}
\label{sec:RelatedWork}
The term ``metaverse" was first used in the science fiction novel ``Snow Crash" in 1992~\cite{stephenson2003snow}. In 2008, Hendaoui \textit{et al.}~\cite{hendaoui20083d} proposed an early concept of the metaverse, i.e., a 3D virtual social world. Schumacher \textit{et al.}~\cite{schumacher2022metaverse} discussed that the metaverse is an immersive Internet which provides a more advanced and efficient platform for social exchange and communication. Up to now, there have been a number of studies that have discussed metaverse from different aspects, as shown in Table \ref{realted work table}.

\textbf{In terms of the concept and definition of metaverse}, Dionisio \textit{et al.}~\cite{dionisio20133d} proposed four characteristics of metaverse, including universality, realism, scalability, and interoperability. Lee \textit{et al.}~\cite{lee2021all} summarized eight basic techniques for building metaverse. Park \textit{et al.}~\cite{ park2022metaverse} discussed three components of the metaverse, namely hardware, software and content. Yang \textit{et al.}~\cite{yang2022fusing} proposed that artificial intelligence and blockchain technology will play a great role in metaverse construction, and further investigated the possibility of integrating artificial intelligence and blockchain technology with the metaverse. The concept of metaverse is now relatively clear, and therefore the number of metaverse applications has started to increase.

\textbf{In terms of metaverse applications}, Chen \textbf{\textit{et al.}~}\cite{chen2022user} explored the advantages of using virtual reality to socialize in a metaverse compared to traditional social media. Tarouco \textit{et al.}~\cite{tarouco2013virtual} proposed a virtual classroom capable of teaching calculus that can help students learn efficiently. Duan\textit{et al.}~\cite{duan2021metaverse} implemented a blockchain-powered metaverse project based on the Chinese University of Hong Kong. Chen \textit{et al.}~\cite{chen2022digital} proposed that applications including healthcare, social software, entertainment, and smart city construction can be implemented in the world of metaverse. In summary, Metaverse applications have exploded in recent years, but there are still some issues with application security.

\textbf{In terms of metaverse security}, Wang \textit{et al.}~\cite{wang2022survey} pointed out that the metaverse is rather vulnerable to problems such as smart contract vulnerabilities, ransomware, scams and phishing. In addition, Lee \textit{et al.}~\cite{lee2021all} proposed that protecting digital assets and user privacy will become an important issue to be addressed. 

Kadar~\cite{Kadar} claimed that in metaverse frauds, there are not only traditional account takeover attacks like phishing scams and hacking, but also account gang money laundering, rug pull and other new attack methods. Kshetri\textit{et al.}~\cite{Nir2022} found that Non-Fungible Tokens, as one of the components of the metaverse, are very vulnerable to hacking. In conclusion, the current security of the metaverse is relatively weak, and there is a large amount of money flowing on it, which can easily attract the attention of criminals.
The security of the metaverse is relatively weak, and there is a large flow of money, easy to attract the attention of criminals. However, the current regulatory strategy on metaverse security is not yet well developed, and accountability for security incidents and crimes is more difficult.

\textbf{In terms of policies related to metaverse}, Ning \textit{et al.}~\cite{ning2021survey} summarized the policies and representative enterprises of metaverse in various countries. Smaili \textit{et al.}~\cite{smaili2022metaverse} argued that laws and regulations should be formulated and updated to protect personal interests and intellectual property rights in the metaverse. Tom \textit{et al.}~\cite {Tom} raised issues such as possible intellectual property rights in the metaverse and suggested how existing laws and regulations can be applied to the metaverse.

Along with the gradual maturation of blockchain and Web3 technologies, the economic system of the metaverse is defined as a digital economic system with decentralized and cross-platform characteristics~\cite{chen2022digital}. The metaverse economy enabled by Web3 has flourished in recent years and has attracted a great deal of attentions from the aspects of research, applications and investments. 
However, this emerging metaverse ecology currently lacks clear regulations and supervision, making it gradually become a hotbed of crimes.
Financial crimes that frequently occur on the metaverse can cause huge economic losses to customers, institutions, and even countries, and affect the long-term health of the metaverse. Therefore, this paper focuses on financial crimes occurring on the metaverse, such as frauds, money laundering, illegal services, code vulnerabilities, etc.


\begin{table}[!ht]
\caption{related work}
\label{realted work table}
    \centering
    \scalebox{0.9}{
    \begin{tabular}{p{18pt}p{120pt}p{62pt}p{15pt}}
    \hline
        \textbf{Refs} & \textbf{Contribution} & \textbf{Keyword} & \textbf{Year} \\ \hline
        \hline
        ~\cite{hendaoui20083d} & Metaverse is a 3D virtual social world. & Metaverse Concept & 2008 \\ \hline
        ~\cite{dionisio20133d} & Proposed four characteristics of metaverse. & Metaverse Concept & 2013 \\ \hline
        ~\cite{lee2021all} & Proposed eight basic techniques for building metaverse. & Metaverse Concept & 2021 \\ \hline
        ~\cite{park2022metaverse} & Discusses the three components of the metaverse. & Metaverse Concept & 2022 \\ \hline
        ~\cite{yang2022fusing} & Proposes the role of artificial intelligence and blockchain technology in the metaverse. & Metaverse Concept & 2022 \\\hline
        ~\cite{schumacher2022metaverse} & Metaverse is immersive internet. & Metaverse Concept & 2022 \\ \hline
        ~\cite{yang2022fusing} & The convergence of artificial intelligence, blockchain technology and metaverse. & Metaverse Concept & 2022 \\ \hline
        
        ~\cite{tarouco2013virtual} & Virtual classroom. & Metaverse Applications & 2013 \\ \hline
        ~\cite{duan2021metaverse} & Virtual university. & Metaverse Applications & 2021 \\ \hline
        ~\cite{chen2022user} & Advantages of metaverse compared to traditional social media. & Metaverse Applications & 2022 \\ \hline
        ~\cite{chen2022digital} & Metaverse applications in healthcare, social software, entertainment, and smart cities. & Metaverse Applications & 2022 \\ \hline
        
        ~\cite{lee2021all} & Protecting digital assets and user privacy. & Metaverse Security & 2021 \\ \hline
        ~\cite{wang2022survey} & Metaverse unavoidable security issues. & Metaverse Security & 2022 \\ \hline
        ~\cite{Kadar} & The existence of traditional and new types of attacks in the metaverse. & Metaverse Security & 2022 \\ \hline
        ~\cite{Nir2022} & Non-homogenized tokens are vulnerable to attacks & Metaverse Security. & 2022 \\ \hline
        
        ~\cite{ning2021survey} & Related policies and corporate layout. & Metaverse Policies, Regulations & 2021 \\ \hline
        ~\cite{smaili2022metaverse} & Protection of personal interests and intellectual property. & Metaverse Policies, Regulations & 2022 \\ \hline
        ~\cite{Tom} & How existing laws apply to the metaverse. & Metaverse Policies, Regulations & 2022 \\ \hline
    \end{tabular}
    }
\end{table}

\begin{figure*}
    \centering
    \includegraphics[width=0.8\linewidth]{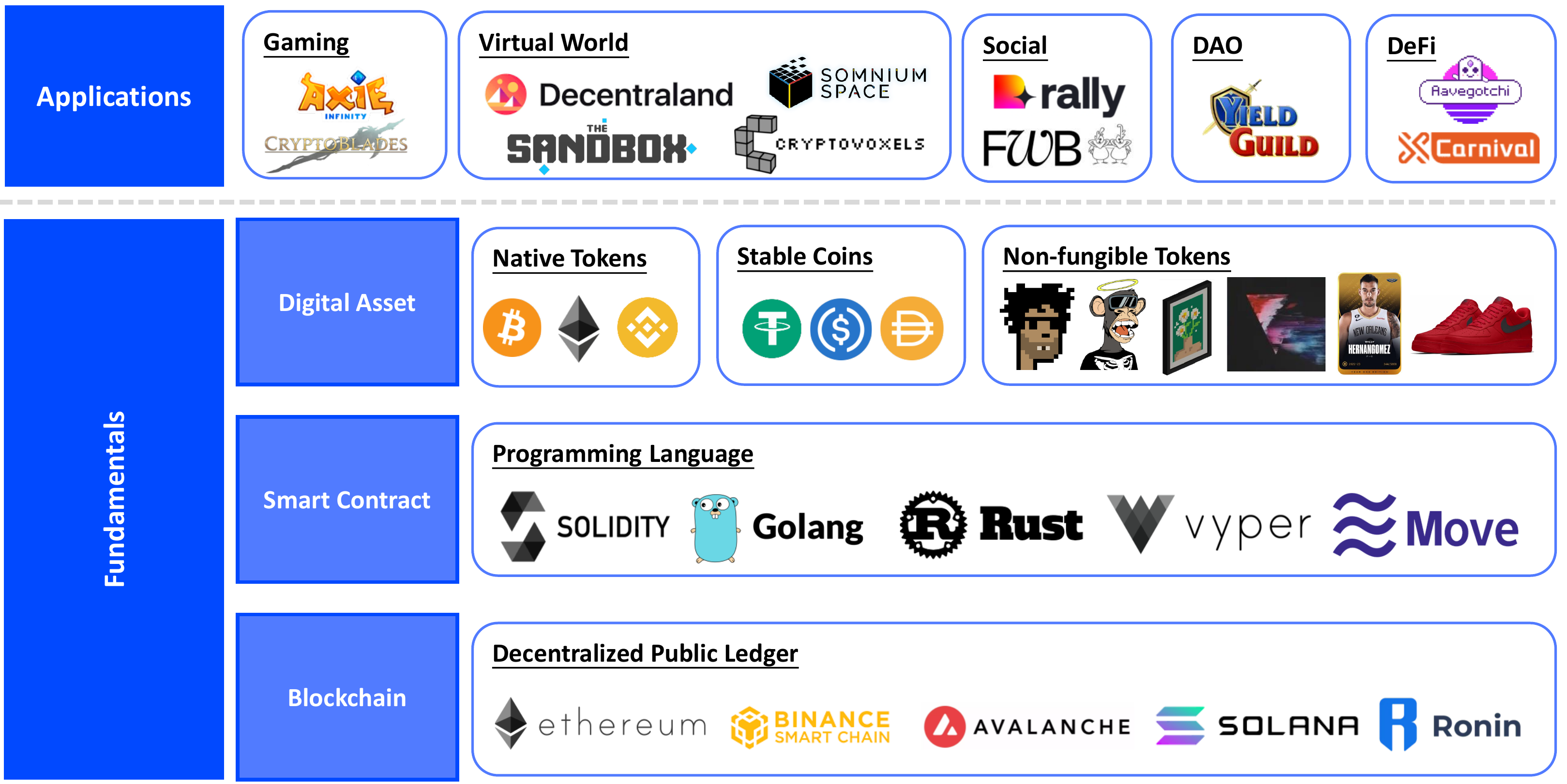}
    \caption{The technical fundamentals of Web3 metaverse economic system and the upper popular applications.}
    \label{fig:fundamental_application}
\end{figure*}

\section{Background and Fundamentals}
\label{sec:BackgroundandFundamentals}
\subsection{Metaverse and Web3}
According to existing studies~\cite{cao2022decentralized}, metaverse and Web3 are closely related concepts. In this section, we will briefly explain Web3, metaverse, and their relationship specifically. In addition, to better understand the ecosystem of the metaverse, we will introduce some key concepts of the metaverse economy, such as blockchain, smart contracts, and crypto assets.

In 1992, when the concept of \emph{metaverse} was first introduced, it was described as a virtual world where people who were not part of the elite could spend most of their free time. In the past decade, the industry is gradually defining the concept and main features of metaverse, and the movie ``Ready Player One'' is regarded as a good interpretation of the metaverse. Currently, the industry considers that blockchain and Web3 are the key technologies supporting the metaverse~\cite{Anu2022web3}.


Since the birth of the World Wide Web in 1989~\cite{gillies2000web}, the Internet has gone through the Web1 era with TCP/IP and other open protocols as the underlying technology. The Web1 era was characterized by a read-only Internet, in which users could click on links on web pages to browse text, images, and other content set by developers, but the web pages could not capture information about the users, and the users could not leave a personal trace on the site. Later, Web1 was replaced by Web2. In the Web 2 era, users could not only browse the Web, but also create their own content and upload it to the web to share with others. Yet Web2 suffers from centralization issues, where user-generated data, such as identity, transaction history, and credit scores, can be easily captured or profited from by companies~\cite{chen2022digital}. Today, the Internet is evolving towards the era of Web3, which is enabled by blockchain and artificial intelligence. Web3 is characterized by decentralization and openness, and the featured Web3 products such as decentralized finance (DeFi), non-homogeneous tokens (NFT), and tokens are necessary components in the metaverse.

\subsection{Fundamentals of Metaverse Economy}
To better understand the economic system of the metaverse, in this section, we first introduce the technologies that implement the underlying logic of the economic system, i.e., blockchain and smart contracts. In addition, the fuel for user activity in the economic system, i.e., crypto assets, also needs to be focused on. The details of the presentation are shown in FIGURE~\ref{fig:fundamental_application}.


\subsubsection{Blockchain}
Anu \textit{et al.}~\cite{Anu2022web3} suggested that in Web3, application data is no longer stored in a private database but in a blockchain that can be written or read by anyone. Blockchain returns digital sovereignty to the users through a decentralized manner. As the underlying ledger of Bitcoin, a blockchain~\cite{nakamoto2008peer} is composed of blocks and chains. User-generated data is stored in a newly generated block that will be linked to previously generated blocks, all of which are strung together in a chronological order~\cite{monrat2019survey}. Each node on the blockchain possesses a complete record of the blockchain data. Therefore, if one node makes an error or is attacked, the remaining millions of nodes can correct the possible errors~\cite{cai2018decentralized,lee2021all}. 
There exist three main types of blockchain: public, private and consortium~\cite{monrat2019survey}, and one of the typical applications of the public blockchain is the Bitcoin.

\subsubsection{Smart Contract}
Szabo introduced the concept of smart contracts in the mid-nineties~\cite{szabo1997formalizing}, where he suggested embedding the logic of the contract into the code. With traditional contracts, a document outlines the terms of the relationship between two parties, which can be enforced by law. If party A violates the contract, party B can take party A to court for non-compliance. A smart contract~\cite{zou2019smart} can be understood as an automatically executed contract with the terms of the agreement between the buyer and seller embedded within the code logic. Languages that currently support writing smart contracts include Solidity, Go, Java, and more.

\subsubsection{Digital Assets and Tokens}
Digital assets are intangible digital objects with verifiable and ownable digital values~\cite {Anu2022web3}. One of the main representatives of digital assets is \emph{token}. A token is a digital asset implemented in a smart contract and is the medium for the storage and exchange of value in the metaverse ~\cite{chen2020understanding}. The benefits of digital assets include a ubiquitous ledger, transparent updates, and payments that can be recorded and verified and do not require centralized settlement~\cite{lyons2022keeps}. In the metaverse, the blockchain automatically records the human interactions in a tamper-proof public ledger, and the block miners obtain tokens as a reward. Tokens can be divided into two types: fungible tokens and non-fungible tokens.
A fungible token is one that is interchangeable with another token while the non-fungible tokens (NFTs) are not interchangeable.

Among various token standards, Ethereum Request for Comment 20 (ERC-20), which was born in 2015, is one of the representative standards for fungible tokens created using the Ethereum blockchain. 
Despite the fact that fungible tokens have the same price per unit of currency, their values are typically somewhat volatile.
For example, Bitcoin (BTC) is approximately 10 times more volatile relative to the U.S. dollar than between major national currencies~\cite{yermack2015bitcoin}. Stable coins were therefore created to address this problem~\cite{chohan2019stable}. Most of these stablecoins solve the problem of the large price fluctuations of most cryptocurrencies denominated in US dollars by pegging their value to a fixed number of traditional monetary instruments~\cite{chohan2019stable}. One representative is USDT, which is pegged to the US dollar.

The functionality of fungible tokens is limited because each coin has the same price and cannot carry extra information. Therefore, the concept of non-fungible tokens is introduced in the ERC-721 standard. Each non-fungible token is unique, with a unique Token ID and a different value. Each non-fungible token tends to be linked to a URL that stores media data, such as images and videos. 
For example, the NFT CryptoPunks series\footnote{\url{https://www.larvalabs.com/cryptopunks}}, which currently has a total value of 1,063,845 Ether on the OpenSea platform, has a historical NFT trading price ranging from 0.001 Ether to several hundred Ether.

\subsection{Applications}
\label{subsec:Applications}
\subsubsection{Game}
The metaverse has the potential to add realism to games. Players are able to build their own territory in metaverse games at will, and are even able to change the pattern and future direction of the games by holding game tokens. The most important feature of  metaverse games is that the currency or circulating items in the games are cryptocurrencies, which means that the currency and items traded in the game are tied to fiat currency. As a result, the concept of ``Play-to-Earn'' is frequently mentioned in metaverse games, that is, earning in-game money in the real world. The Play-to-Earn model was started with a pet trading and fighting game called Axie Infinity \footnote{\url{https://whitepaper.axieinfinity.com/}}. Players can create virtual pets called ``Axies" in this ethereum-based game, and then send the avatars to earn game tokens through competitions and battles. There is also CryptoBlade \footnote{\url{https://www.cryptoblades.io/}}, a game based on the Coinan chain. In the game, users can use weapons to defeat enemies in the game to earn Skill tokens, which in turn upgrade the character. As users level up in the game, they can make weapons with different power. The more powerful the weapons they have, the more Skill tokens they will get.

\subsubsection{Virtual World}
Typical examples of virtual worlds are Decentraland \footnote{\url{https://decentraland.org/}}, Cryptovoxels\footnote{\url{https://www.voxels.com/}}, The Sanbox \footnote{\url{https://www.sandbox.game/en/}} and Somnium Space \footnote{\url{https://somniumspace.com/}}. In the virtual world, users can own a part of the world by purchasing scarce land NFTs. Since the land in the world is represented by NFT, this means that each plot is unique and ownership can be easily tracked. In addition, since the land within the platform acts as a virtual asset, it supports the owner to build it at will. Therefore many users will move real-life buildings into these platforms, such as casinos, financial centers and shopping malls.

\subsubsection{Social}
Social projects related to metaverse concept are mainly specialized social applications for certain kinds of groups, such as Friends with Benefits\footnote{\url{https://www.fwb.help/}}, and Rally \footnote{\url{https://rally.io/}}. For example, Friends with Benefits is a Discord-based private club with over 2,000 members, which requires not only a strict identity check (written application), but also nearly 10,000 dollars in tickets and a certain amount of the native tokens called FWB to build their own customized personal tokens on the platform. Rally is a incentive platform for creators and their communities to build their own independent digital economies. Creators receive incentives for creating personal tokens, and fans can support their favorite creators by purchasing tokens with tokens RLY.

\subsubsection{DAOs}
Decentralized autonomous organizations (DAOs) are groups established on a mission to coordinate and collaborate through a set of shared rules implemented on the blockchain. For example, Yield Guild Games \footnote{\url{https://yieldguild.io/}}is a DAO and gaming guild founded in the Philippines and composed of players and investors. In Yield Guild Games, all participating members are investors and players. The investors are responsible for the NFT assets of the different games in their ecosystem and lend these assets to the community of gamers. Players are responsible for participating in the game in order to acquire game currency, while the guild can reinvest the available surplus in order to purchase virtual assets and land within the game.

\subsubsection{DeFi}
DeFi was introduced in 2020, and its rapid development in recent years has greatly boosted the demand for NFT. 
Digital art, domain names, and anything related to ownership can become NFTs. Like art and other collectibles, these digital assets are more difficult to buy and sell due to their more specific markets. However, the advent of DeFi has changed all that, bringing much-needed liquidity to NFTs. DeFi has flourished in recent years because it has democratized access to financial services such as lending, savings and insurance, which has attracted significant investment into the DeFi ecosystem. In short, the emergence of DeFi is revolutionizing the traditional financial industry, and a large number of influential financial practitioners are paying attention and getting involved.

XCarinval \footnote{\url{https://xcarnival.fi/}}, for example, is a collateralized lending platform for metaverse assets, offering collateralized lending services including all types of NFT assets as well as long-tail assets, providing an effective value release for liquidity-starved assets. In addition, Aavagotchi \footnote{\url{https://www.aavegotchi.com/}} is a digital collection of NFTs combined with DeFi, where each Aavegotchi NFT token is both a collectible and an asset capable of generating income, and the properties of the NFTs depend on their value and rarity in the Aavegotchi universe.

\section{Financial Crimes in Metaverse}
\label{sec:Crimes}

Financial crimes are often defined as crimes against property and involve the unlawful transfer of money or other types of property belonging to another person. In the process of committing financial crimes, offenders usually use illegally acquired property for personal use and benefit. According to the International Monetary Fund, financial crime is ``any non-violent crime that generally results in a financial loss"~\cite{IMF2001}. In the UK, the Financial Services and Markets Act defines financial crime as ``any offence involving fraud or dishonesty; misconduct in, or misuse of information relating to, a financial market; or handling the proceeds of crime"~\cite{FSMA2000}. Typical financial crimes include scams, wash trading, money laundering, etc. These crimes not only bring losses to investors and users, but also pose a certain degree of threat and challenge to the current economic ecology.

Along with the recent development of Web3, financial crimes have been given a more diverse and complex meaning in the metaverse ecology. 
Based on the employment of blockchain in metaverse, many fraudsters have found new opportunities for illicit profits, including money laundering, identity theft, and scams~\cite{Kadar}. 
Due to the decentralized and anonymous nature of the metaverse, financial crimes on the metaverse are usually more covertly disguised, making crime detection and financial regulation on the metaverse more difficult.

Therefore, while the advent of the metaverse era has injected new momentum into the financial system and created trade opportunities for social commerce, the lack of effective regulation on blockchain or Web3 may make the metaverse a hotbed of criminal activities, promoting the occurrence of financial crimes such as scams, code exploits, wash trading, money laundering, and illegal services and shops. To this end, a summative research work on metaverse financial crimes is particularly urgent and critical. In this section, we provide an overview and taxonomy of financial crimes that have emerged since the development of the metaverse.

    

\begin{figure*}
    \centering
    \includegraphics[width=0.95\linewidth]{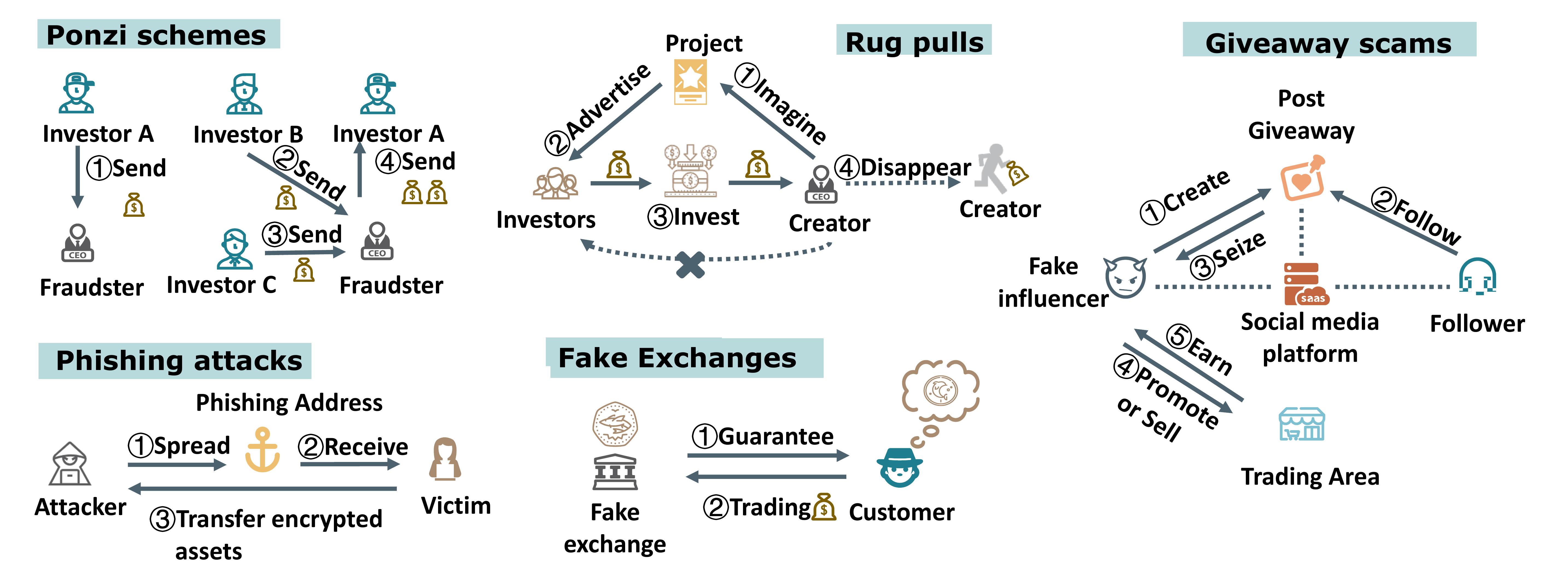}
    \caption{The schematic diagram of various types of scams in the metaverse.}
    \label{fig:types_of_scams}
\end{figure*}

\subsection{Scams}
The concept of scam has been around for a long time, with the so-called Golden Age of the Great Scam having been documented in the relevant academic literature in the late 19th and early 20th centuries~\cite{lindesmith1940big,brannon1948game}. Subsequently, Market Business News  defined scam as a dishonest or fraudulent scheme. Such a scheme attempts to obtain money or something of value from people and is a confidence trick perpetrated by a dishonest group, individual or company. The person or organization committing the scam is often referred to as a scammer, trickster, or swindler~\cite{MBN2022}. 
Whereas scams used to occur frequently in offline social interactions, with the growth of the Internet in the past decades, scams have successfully infiltrated online networks.
While scams used to occur frequently in offline social interactions, in recent years, with the growth of the Internet, scams have successfully infiltrated online networks. More specifically, in the cryptocurrency market, scammers use the pseudonymous characteristics of cryptocurrencies to perpetrate untraceable crypto-asset scams and attempt to defraud investors for ill-gotten gains~\cite{bartoletti2021cryptocurrency}. 
Scams on meterverse can be categorized into the following types~\cite{bartoletti2021cryptocurrency,elliptic,APWG2021}: (i) Ponzi schemes; (ii) Rug pulls; (iii) Phishing attacks; (iv) Fake exchanges; and (v) Giveaway scams. The basic workflows of various types of scams in the metaverse are shown in FIGURE~\ref{fig:types_of_scams}.

Based on the discussion in Section III, the metaverse needs to rely on the existence of crypto markets to operate. This fact has led, on the one hand, to an increasing amount of retail and institutional capital entering the crypto space, bringing a wealth of crypto assets to the metaverse; on the other hand, investors in the emerging economic ecosystem of the metaverse usually have limited education in keeping their money safe and identifying illegal activities, etc~\cite{elliptic}. As a result, a steady stream of scams has also emerged in the metaverse. Furthermore, the diversity of crypto assets in the metaverse has allowed for an expansion of the types of scams. Derivative scams such as Fake metaverses, Fake land expansions, Technical support scams, and 3D social engineering have emerged (these derivative scams will be discussed in detail in the Emerging Crimes in the Metaverse section)~\cite{elliptic}. With the development of the metaverse, scams have gradually become a major concern for the security of the decentralized financial system in the metaverse.

\subsubsection{Ponzi schemes}
Back in December 2017, the staff at Sky Mavis, a Vietnamese studio, conceived the idea of combining cryptocurrency with online gaming, and set out to create an Ethereum-based online game called Axie Infinity~\cite{S.Mavis}. This game allows players to own virtual assets and rewards those who are able to reach advanced skill levels. However, due to the introduction of the Play-to-Earn (P2E) concept in the game~\cite{alam2022understanding}, many players from low-income South Asian countries began to use the game as their main source of income~\cite{J.Brustein,K.Servando}. This phenomenon has attracted wide attention, and opponents have begun to question this Ponzi-like mechanism of making money from games.


\subsubsection{Rug pulls}
In early 2022, the creators of the ``Big Daddy Ape Club" project, associated with the NFT private collection, presented a private collection of 2222 apes to investors. It then received an investment amount of 9,136 SOL, or about \$1.3 million, before disappearing from the public eye. This was called the largest ``Rug pulls" in the history of the Solana blockchain~\cite{coincu2022}. Notably, it had been verified by decentralized identity verification company Civic before that. However, this fraudulent act of raising funds and breaking promises still took place.


\subsubsection{Phishing attacks}
In the same year, Sandbox Game, a community-driven platform that provides a service for user creators to create and monetize voxel assets and gaming experiences on the blockchain, has received the onslaught of fraudulent websites claiming to be releasing land for sale in the second quarter~\cite{reddit2022}. These fraudulent sites have designed and disguised their pages to launch phishing attacks on users of The Sandbox Game.


\subsubsection{Fake exchanges}
Fake exchanges are also a type of scam that may pose a threat to the security of the metaverse. In December 2017, the Bitcoin community and South Korean authorities exposed a fake exchange called BitKRX~\cite{cointelegraph2017}. BitKRX lured innocent users by posing as a legitimate exchange and disguising itself as an offshoot of the large and reputable trading platform KRX to scam users out of the amount of money they were holding.


\subsubsection{Giveaway scams}
Recently, some cases of Giveaway scams related to metaverse assets have gradually emerged. In March 2022, when the Yuga Labs team announced the launch of MetaRPG and the native crypto asset ApeCoin, a number of fraudulent actors on social media platforms tried to trick users into clicking on malicious links or sending funds for fraudulent giveaways. In the process, these fraudulent actors managed to raise approximately \$900,000~\cite{elliptic}.



Researchers have begun to explore the crypto scams involved in the metaverse. Bartoletti \textit{et al.}~\cite{bartoletti2021cryptocurrency} review the scientific literature on cryptocurrency scams. A systematic classification of scams was performed based on a new taxonomy and a unified dataset consisting of thousands of cryptocurrency scams was created by collecting data from different public sources. 
Smaili \textit{et al.}~\cite{smaili2022metaverse} provide strategies to deter, detect, and prevent scams by considering both individual (user) and organizational levels. Currently, research on metaverse frauds is still in a relatively preliminary stage, while in-depth methods and applications need to be further explored.

\subsection{Code Exploit}
Since the success of Bitcoin, the applications of blockchain technology gradually emerging in many fields and services, such as financial markets, Internet of Things, supply chain, healthcare and storage. Since these systems usually store rich information,  blockchain has also become a high-value target for cybercriminals or hackers~\cite{lin2017survey}. Such attacks on the blockchain are usually manifested in various ways of ``hacking" into the blockchain system. For example, some blockchain attacks focus on the poor protection of private keys by blockchain account owners or cryptocurrency exchanges to steal cryptocurrencies or personal assets of others. Other cyber attackers exploit vulnerabilities in blockchain protocols or their smart contract implementations to compromise blockchain systems~\cite{sayeed2019assessing,parizi2018smart}. Attacks in this manner are often referred to as code exploit attacks. On the one hand, protocol design vulnerabilities occur when blockchain architects fail to adequately consider the impact of features built into their technology. A typical example is the attack on the Verge cryptocurrency, whose attackers exploit approximately 10\% of the hashes of the blockchain for a 51\% attack~\cite{sayeed2019assessing}. This problem is not caused by any programming error, but by the design of the protocol itself. On the other hand, many hackers have exploited the vulnerabilities in smart contracts to steal crypto assets. For example, in June 2016, the smart contract code vulnerability of the DAO was maliciously exploited by hackers to empty more than 2 million (\$40 million) Ether coins~\cite{Sirer2016}.

\begin{figure}[htbp]
\centering
\includegraphics[height=3.7cm,width=8.0cm]{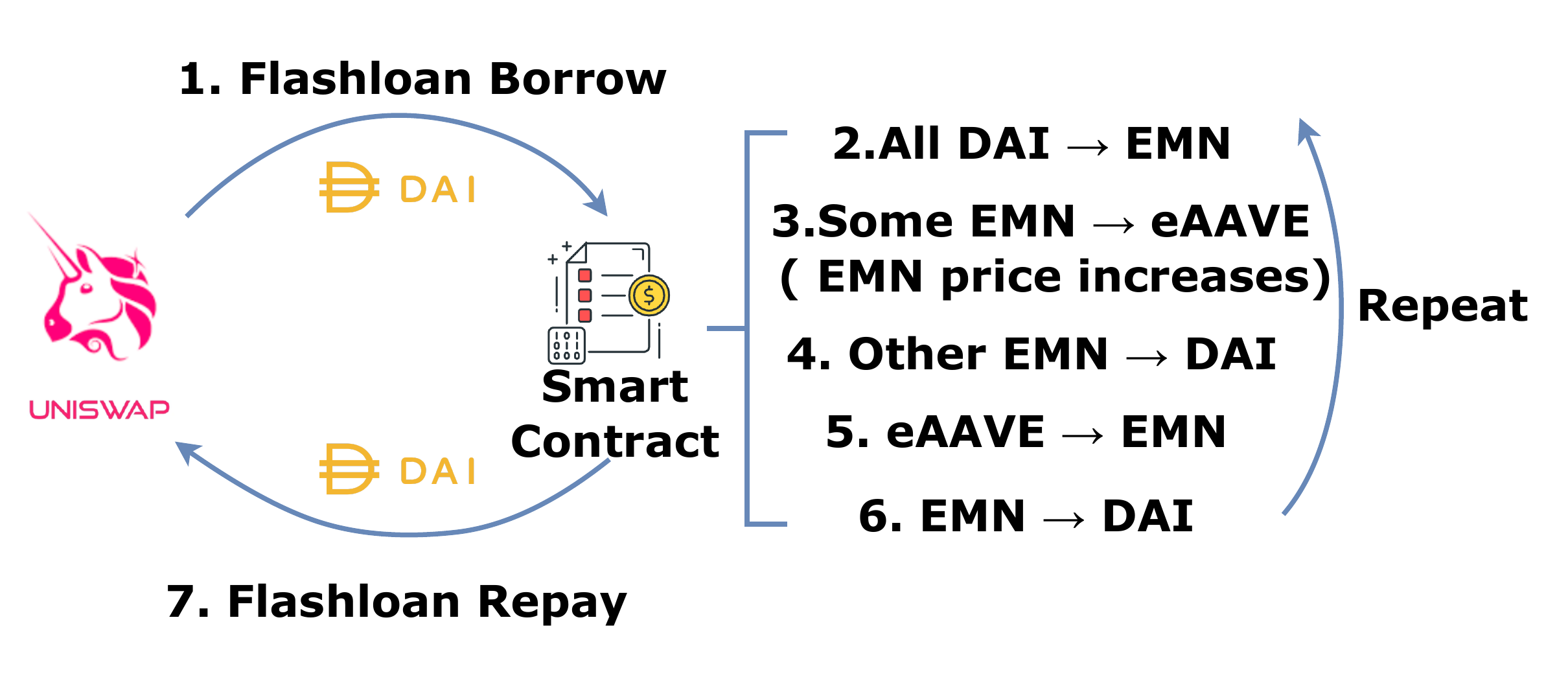}
\caption{Flowchart of a hacker attack on Eminence (EMN) by flashloan.}
\label{EMN}
\end{figure}

Similar to DeFi, there exist a large number of smart contracts in the metaverse. Therefore, cybercriminals can likewise take advantage of poorly structured smart contracts or vulnerabilities in smart contracts to steal cryptocurrencies and NFTs in the metaverse .
There are already examples of code being maliciously exploited in metaverse projects. 
In September 2020, a developer of Yearn Finance developed an NFT game called Eminence Finance, which has its own token called EMN. Without much understanding of this project, some investors discovered the token and minted \$15 million worth of EMN in a few hours. They used a smart contract designed to allow players to exchange DAI (a stable coin) for EMN to fund in-game purchases. However, a hacker discovered a way to deplete the funds in the contract using a lightning loan, which caused the price of the token to drop dramatically. The whole process of the flashloan case on Eminence is shown in Figure~\ref{EMN}. 
On August 21, 2021, some hackers managed to break into the NFT mint contract for the NFT Gods metaverse game protocol on the Binance Smart Chain. They bypassed the private key check through technical means and stole nearly 9 million LG tokens, the platform's native asset. They then sold these tokens for \$1.45 million through a crypto asset exchange service which did not require KYC checks~\cite{elliptic}.Additionally, while NFTs are blockchain-based, exchanges and marketplaces such as OpenSea and Rarible operate in a centralized manner, making them unable to take advantage of peer review systems that can identify and fix errors. As a result, they are vulnerable to code exploit attacks. In September 2021, 42 NFTs worth over \$100,000 disappeared into thin air due to a vulnerability in the OpenSea token marketplace~\cite{btcpeers2021}. 

At this stage, research on code exploit attack has mainly focused on the exploration of smart contract vulnerabilities in Ethereum. However, with the development of the metaverse, research on smart contract vulnerabilities in the metaverse is gradually emerging. Ndiaye \textit{et al.}~\cite{ndiaye2021cryptocurrency} summarized cryptocurrency crimes by assessing the cost of attacks and losses caused by smart contracts. Moreover, they provide an in-depth analysis of the root causes and consequences behind the attacks and the defense strategies that exist. Kshetri \textit{et al.}~\cite{Nir2022} discussed malicious attacks and disruptions on cryptocurrencies and NFTs in the metaverse, and provided an in-depth analysis of cyber attacks on crypto assets.

\subsection{Wash Trading}
Wash trading is a market manipulation behavior that has appeared in traditional financial scenarios~\cite{Cao2014} and is recognized as a financial crime in most countries. Generally speaking, it refers to the repeated trading of assets in order to provide misleading information to the market. For the purpose of commercial competition or commercial interests, wash trading is generally done by several users who collude with each other and appear to be trading, but in fact they do not change their position or assume any real market risks~\cite{victor2021detecting}. The US Commodity Futures Trading Commission defines wash trading as ``trading or intending to trade that makes the trade appear to be completed without creating market risk or changing the trader's market position''~\cite{CFTCwashtrading}. Wash trading activities inevitably lead to an increase in (fake) trading volume and create a false sense of prosperity.

Wash trading in the metaverse economic system exists mainly in the native cryptocurrency, ERC20 token market and NFT market.
In fact, many exchanges have been accused of inflating trading volumes through wash trading. In March 2019, Bitwise Asset Management reported to the U.S. Securities and Exchange Commission that 95\% of Bitcoin trading volume is fake. By faking volume, the exchanges with the highest trading volumes can receive listing fees from ICOs, reportedly in the millions of dollars~\cite{Bitwise2019}.  In August 2020, Coinbit, third largest cryptocurrency exchange of South Korea, was charged by the police with allegedly faking more than 99\% of its trading volume~\cite{chen2022cryptocurrency}. In the NFT market of the metaverse, wash trading is also quite rampant. 
According to Elliptic~\cite{elliptic}, 95\% of all activities on the decentralized NFT trading platform LooksRare is associated with wash trading. There are two main scenarios of NFT wash trading observed so far. One is the fictitious trading volume in order to get on the new NFT collection, which is similar to ICOs' conducting token washing in order to go public. One of the requirements for the centralized NFT trading platform OpenSea to validate an NFT collection is at least 100 ETH transaction volume, which may be difficult to meet for newly launched collections. This requirement appears to encourage fraudulent transactions, where fictitious transactions are executed between multiple accounts under the control of the attackers to artificially inflate transaction volumes. The other main scenario of NFT wash trading is to obtain other token rewards by wash trading through NFT transactions. Chainalysis~\cite{chainalysisNFT}  reported blatant double trading of three identical NFTs between two wallets, trading approximately 650,000 ETH and costing \$114 million in transaction fees. They ended up with approximately \$185.5 million worth of tokens from the NFT trading platform, bringing in nearly \$71 million in profits.
This type of wash trading is not without its victims. First, the metaverse project trading platform pay rewards for trading in the fake NFT trading activity, and the wash traders illegally take the rewards of the NFT trading. Second, NFT collectors throughout the metaverse marketplace may be misled by the wash trading into believing that there was active trading activity in native currency or the marketplace for wearables and land NFTs of this metaverse. Therefore, wash traders influence the valuation of metaverse assets and even manipulate the metaverse market.

It is extremely challenging to detect and count wash trading. The current main methods of wash trading are based on transaction data. Some work constructs a transaction network by abstracting the transfer relationship between users and employs the network topology theory to analyze and detect the wash trading activities. For example, Victor\textit{et al.}~\cite{victor2021detecting} proposed a wash trading detection method for decentralized exchanges based on the identification of network loops and cycles. This method found various wash trading structures and manipulated volumes of IDEX and EtherDelta with a total value of \$159 million. Serneels \textit{et al.}~\cite{Serneels2022} proposed three methods to flag suspicious NFT wash trading activities, including closed loop token trades, closed loop value trades, and high transaction volumes. Inspired by Victor \textit{et al.}~\cite{victor2021detecting}, \cite{das2021understanding} detected suspicious wash trading in NFT and counts the proportion of suspicious wash trading in the NFT set to the total transaction volume, as well as the proportion of wash trading in the mainstream NFT trading market. Existing work also proposes to design washing behavior indicators through empirical analysis. For example, Chen \textit{et al.}~\cite{chen2022cryptocurrency} designed several metrics to analyze wash trading on centralized exchanges based on off-chain transaction data and on-chain transaction data.

\subsection{Money Laundering}
\begin{figure}[htbp]
\centering
\includegraphics[height=5.5cm,width=8.0cm]{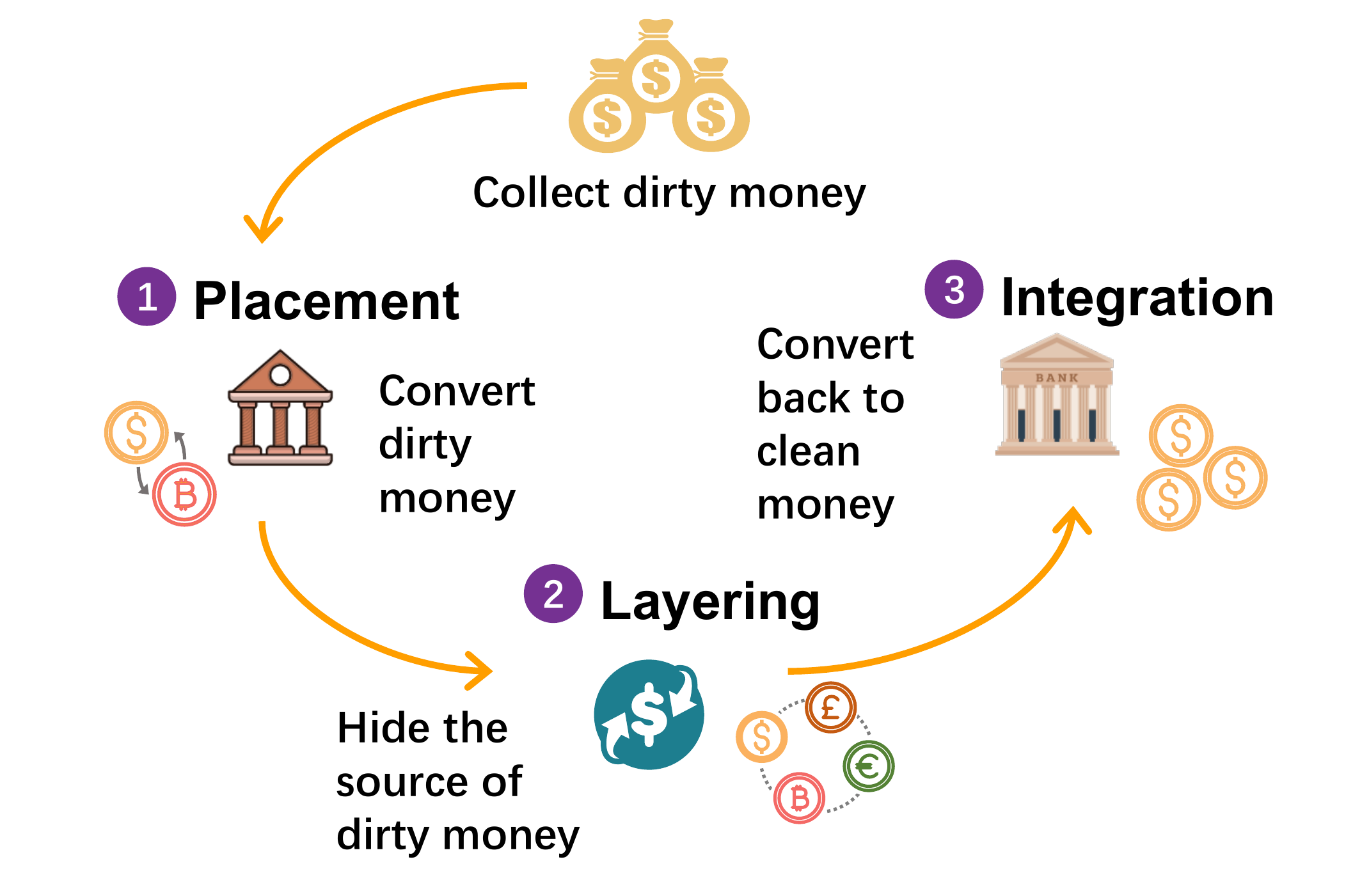}
\caption{The three specific steps of Money Laundering}
\label{fig:money_laundering}
\end{figure}

Money laundering, which is a serious financial crime that fuels crimes such as drug trafficking and terrorism, has a negative impact on the global economy. The Association of Internationally Recognised Anti-Money Launderers (ACAMS) defines money laundering as acquiring the proceeds of crime and disguising their illicit origin in order to use those funds for legal or illegal activities~\cite{ACAMS2012}. Intuitively, money laundering is the process of making dirty money look clean. The process of money laundering can be subdivided into three specific steps, as illustrated in FIGURE~\ref{fig:money_laundering}, namely, placement, layering, and integration~\cite{FATF2021}. First, illicit funds are surreptitiously channeled into the legitimate financial system. Then, complex financial transactions are used to hide the source of illicit funds, sometimes by wire transfer or by transferring money through numerous accounts to create confusion.
Finally, the funds are integrated into the financial system through additional transactions until the  ``dirty money'' looks ``clean''. Due to the huge negative financial impacts of money laundering crimes on society, most financial organizations now have anti-money laundering (AML) policies in place to detect and prevent such activities~\cite{FINRA2021}. 

In a metaverse ecosystem where crypto assets such as NFsT are widely used but legal regulation of metaverse transactions is still immature\cite{hartwich2022probably}, the potential for money laundering is considered high. In particular, with the growing trend of total sales of land assets and wearables (also known as ``skins'', which are clothing and accessories for avatars) in the metaverse, it is highly likely that criminals will use these new assets for illegal money laundering operations.

In terms of metaverse land assets, the total sales of all crypto assets (including land assets in virtual platforms) in the blockchain-powered virtual reality platforms like Decentraland, Cryptovoxels, the Sandbox, and Somnium Space have exceeded \$500 million in 2021~\cite{elliptic}. Of those, with millions of dollars of plots sold, the average land value in Decentraland reaches tens of thousands of dollars by 2021~\cite{elliptic}, indicating a potential way for a large amount of illicit money to be transferred. In addition, unlike real world purchases of property or land, purchasing metaverse land often requires only a crypto asset address and some funds without KYC checks, which also makes it extremely convenient for money laundering and other criminal activities.
In terms of wearable device assets, the wearable market is expected to reach \$3 trillion by the end of 2023. Criminals can buy a wearable device in one metaverse and then move it to another, cashing out through secondary sales, thus making the money flow harder to track as it spans multiple blockchains. These data suggest that as metaverse financial assets continue to evolve, illegal actors are likely to use them as a conduit for laundering illicit assets that may come from real-world activities or other crypto-based crimes, and criminals can hide their origin~\cite{elliptic} by exchanging them for metaverse-based assets (e.g., land in the metaverse, wearable devices, etc.).
These data suggest that as metaverse financial assets continue to evolve, it is likely that illicit actors will use them as a primary conduit for laundering illicit assets that may come from real-world activities or other crypto-based crimes, and that criminals can hide their origin by exchanging them for metaverse-based assets (e.g. land in a metaverse, wearables, etc.)~\cite{elliptic}.

With the booming growth of the metaverse and the expansion of crypto assets therein, it is urgent to conduct the investigation and prevention of money laundering crimes on the metaverse economic ecosystem. Recently, Qin \textit{et al.}~\cite{qin2022identity} discussed money laundering crimes in the crypto market and analyse the legal level measures proposed by countries or regions such as the EU, Japan, and the US on the prevention of cryptocurrency money laundering crimes in the context of the metaverse. Moreover, Gu et.al~\textit{et al.}~\cite{gu2022chain} and Kampers et al.~\cite{kampers2022manipulation} proposed algorithms to detect unusual transactions in cryptocurrencies in 2022.

\subsection{Emerging Crimes in Metaverse}
\begin{figure}[htbp]
\centering
\includegraphics[height=5.0cm,width=8.0cm]{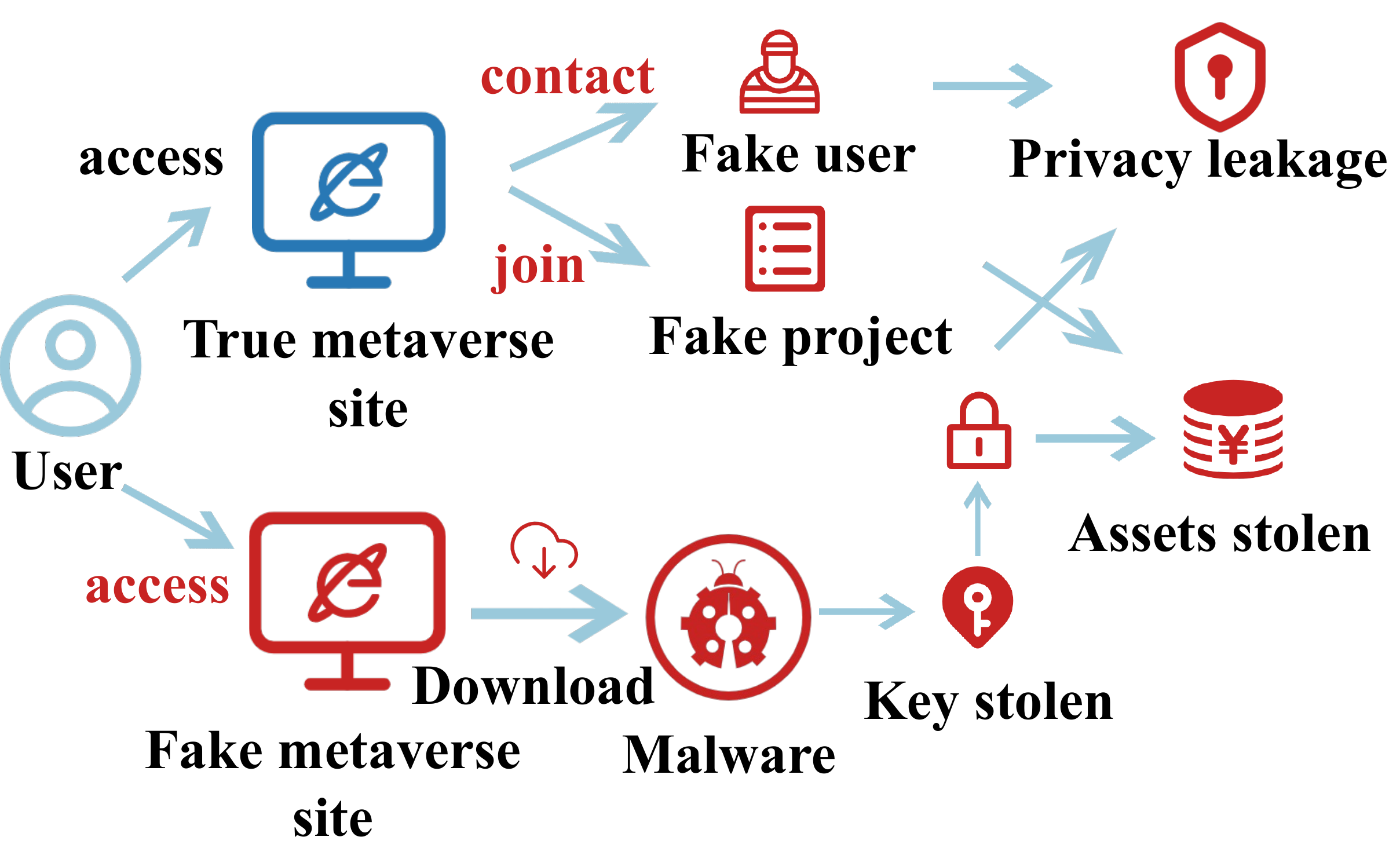}
\caption{Common fraud pitfalls in metaverse}
\label{emerging}
\end{figure}

Financial crimes have existed for centuries, but in the metaverse, crimes have taken on a more multifaceted meaning, with multiple types of crime closely related to metaverse crypto assets.
Figure~\ref{emerging} shows the trap that users may fall into. In this section, six emerging and noteworthy forms of crime, namely, illegal services and shops, fake metaverses, fake land expansions, technical support scams, 3D social engineering, and sanctions and terrorism funding, will be discussed in turn.

\subsubsection{Illegal Services and Shops}
Art galleries, topic-based museums, and project social spaces are popular applications in the metaverse which are closely related with their corresponding real-life scenes. Hybrid shops are a typical way to combine virtual and reality in metaverse business. These shops allow users to purchase items in a metaverse and then pick up the goods offline. During the Decentraland Fashion Week in March 2022, there are shops selling clothing and accessories that allow users to buy goods in a metaverse and have them delivered to a realistic address provided by the user. However, unscrupulous actors may use these shops to sell illegal services and goods, using the items in metaverse shops as a cover for illegal goods in the real world. However, instead of seeing metaverse shops as their main expansion opportunity at the moment~\cite{elliptic}, illicit actors seem to be focusing their attention on other dark web activities.

\subsubsection{Fake Metaverses}
Criminals sometimes defraud users of their account information and crypto assets by announcing the launch of a new metaverse project or by creating a fake NFT website that resembles a legitimate project. In the case of a fake metaverse, illicit actors may use social media and marketing promote a metaverse which does not exist. When users try to connect through their MetaMask they see their accounts wiped out. In the scenario of fake NFT website, the user is taken to a fake website that appears to be the same as the legitimate website by means of a link, and is directed to download the malware present on the website to the user's device. 
Through malware intrusion, the stored credentials of the user are stolen and used to access the cryptocurrency wallet of the user and transfer all of the crypto assets of the user. A case in point is the fake clone of the MetaversePRO website~\cite{trendmicro2022}, a fair, community-managed Metaverse financial infrastructure platform, where the fake clone of the website was designed to be indistinguishable from the legitimate website by choosing a slightly different domain name and visually similar pages to make it difficult for users to tell the difference, and then a step-by-step guide to defraud users of their crypto assets.

\subsubsection{Fake Land Expansions}
In a metaverse where the supply of land assets is limited, illicit actors deceive users into buying by launching fake land expansions of new areas or creating fake versions of land assets of famous metaverse projects. This type of scam exists because land resources in the metaverse world are often limited.
For example, the metaverse project of Decentraland started with 90,601 land plots, a portion of which was reserved for community ownership~\cite{elliptic}. As limited land resources are gradually sold, reserved land becomes more expensive and appealing as demand outstrips supply, so illegal actors use the scarcity of land and users' inability to recognize the authenticity of metaverse land to falsify and defraud users.

\subsubsection{Technical Support Scams}
Criminals attempt to induce new users to share their private keys or connect their MetaMask wallets to illegal and malicious websites by impersonating staff or technicians of the Metaverse project
In fact, this type of crime is widespread in many areas related to new technologies. Broadly speaking, this type of crime involves scammers using scare tactics to induce users to pay for unnecessary technical support services~\cite{microsoft2022}.

\subsubsection{3D Social Engineering}
Social engineering attacks are increasingly rampant in today's networks. Attackers aim to manipulate individuals and businesses to leak valuable and sensitive data \cite{kalnicnvs2017security}, using malicious activities accomplished in interpersonal interactions to psychologically influence a person to leak confidential information or compromise security procedures~\cite{pokrovskaia2017social}. This attack is manifested in the metaverse primarily in the form of a 3D avatar that the illicit actor may take to impersonate a colleague of the user to psychologically influence and actively share sensitive information and access through legitimate communication

\subsubsection{Sanctions and Terrorism Funding}
Cyber criminals use crypto assets in the metaverse to assist bad actors to evade sanctions or fund terrorism.
This type of crime intends to conduct illegal fundraising through metaverse-related assets. Although, the amount of most crypto assets is currently relatively small compared to the total amount needed for terrorism activities, somewhat suggesting that the potential for fundraising in this manner is limited. However, as the metaverse ecosystem evolves, this type of crime may become a serious issue of concern. In this regard, an account with a possible link to the Ryuk ransomware has raised suspicions~\cite{elliptic}. Ryuk, a ransomware attributed to the hacker group WIZARD SPIDER~\cite{Avast2022}, was recently discovered. Ryuk is typically used by hackers to target high-value targets, infect systems, and encrypt files in order to force the target to pay a ransom. This account holds NFTs associated with the metaverse and interacts with Ethereum and some ERC20 tokens, and it is estimated that this scammer may have netted a total of \$150 million by the end of 2020. 
While Ryuk and other ransomware campaigns are not sanctioned, the fact that Ryuk and other ransomware campaigns are often operated by sanctioned entities and nation states suggests that the account is likely committing illegal asset fundraising.

\section{Anti-Crimes in Metaverse}
\label{sec:Anti-crimes}
No matter how much emphasis is placed on the ``decentralisation'' and ``data freedom'' of the Web3 metaverse, the issue of regulation is always an key part of the metaverse architecture. In this section, we discuss the current research on the regulation of the Web3 metaverse and its basic components from the perspectives of academic research and related policies and measures.

\subsection{Academic Researches}
With the rise of the metaverse and the widespread occurrence of related financial crimes, some efforts have been devoted to combating these crimes. The following is a brief overview of existing research on anti-crimes in metaverse from the perspectives of three academic disciplines.

\subsubsection{Computer Science Field} 

In the past decade, a series of studies from computer science or software engineering fields focused on blockchain smart contract security, behavioural mining, and anomaly detection. 

In terms of smart contract security, Atzei et al~\cite {atzei2017survey} analysed the security vulnerabilities of smart contracts on Ether, revealing the financial security issues they can cause. In 2022, Kushwaha et al~\cite{kushwaha2022systematic} conducted a systematic review of research on smart contract security issues up to 2022. In addition, security tools~\cite{tsankov2018securify} and analytical frameworks~\cite{brent2018vandal} have been put forward to address security concerns in smart contract. 

In terms of behavioral mining and anomaly detection, according to an overview given in~\cite{wu2021analysis}, existing work can be divided into four parts: entity identification, transaction pattern recognition, illegal activity detection, and transaction tracking. For instance, Victor et al.~\cite{victor2020address} proposed a clustering heuristic for entity identification based on the Ethernet account model, Huang et al.~\cite{Huang2022Ethereum} modeled the Ethereum transaction records as a large-scale transaction network and proposed a GCN-based model to classify account in Ethereum, and Liu et al.~\cite{Liu2022fa} proposed a method called FA-GNN to deal with the heterophily issue for account classification in Ethereum. In~\cite{wu2021detecting}, Wu et al. proposed temporal attribute heterogeneous modalities, and implemented transactional pattern recognition using modal detection. In an environment where the anonymity of cryptocurrencies has led to their widespread use in financial crimes, Akcora et al~\cite{akcora2019bitcoinheist} propose a cryptocurrency-based ransomware detection framework that can be used to automatically detect ransomware. In addition, a series of data modeling and transaction tracking methods have been proposed~\cite{phetsouvanh2018egret,yousaf2019tracing,Lin2021Evolution,lin2020t,lin2020modeling,Jin2022heterogeneous,lin2022ethereum,Zhou2022behavior}.

Recently, some work started to analyze metaverse security. For example, Kshetri et al~\cite{Nir2022} discussed the impact of possible attacks and various types of frauds on NFT. However, anti-crime research on the metaverse is still at a more preliminary stage compared to related research in the blockchain and cryptocurrency.

Outside of academia, the industry has also paid much attention to security issues in blockchain, Web3, and the metaverse. Several cryptocurrency and Web3  services companies have released reports on security and anti-crime issues. For example, CERTIK published HACK3D: The Web3 Security Quarterly Report~\cite{HACK3D}, in the second quarter of 2022. This report states that the security of individual projects in Web3 is dependent on the security of the entire ecosystem; ELLIPTIC analysed potential metaverse financial crime types and proposes corresponding measures to prevent them in a report entitled The Future of Financial Crime in the Metaverse ~\cite{elliptic} published in 2022; SlowMist analysed some typical security incidents and published an advanced analysis method for the tracking of coin blender funds in its Blockchain Security and Anti-Money Laundering Analysis Report for the first half of 2022~\cite{2022SlowMist}. Kaggle, the renowned data modelling and analytics competition platform, is also organising an It's Time to Protect Web3 themed competition in 2022 in partnership with Forta, a community-based decentralised security platform, to identify phishing scam accounts and maintain the security of the Web3 ecosystem.

\subsubsection{Financial Field}
The virtual economic system is a crucial part of the metaverse and the financial community has long studied financial issues in the virtual economy. Smaili et al.~\cite{smaili2022metaverse} flagged the different kinds of fraud risks that can be posed by the metaverse. Wronka et al.~\cite{wronka2021financial} analysed the impact of DeFi on efforts to combat financial crime. Back in 2018, the National Bureau of Economic Research released a study on the Bitcoin economic system ~\cite{abadi2018blockchain}. The Financial Action Task Force on Money Laundering (FAFT), one of the world's foremost international organisations combating money laundering, updated its guidance on virtual assets and virtual asset service providers ~\cite{force2019guidance} in 2021, further requiring countries to assess and mitigate the risks of their virtual asset financial activities. There are corresponding studies in academia, such as Barone et al ~\cite{barone2019cryptocurrency} comparing usury in traditional economic systems with cryptocurrency as a means of money laundering.

\subsubsection{Legal Field}
In the face of the fertile ground that the ecology of the metaverse presents for financial crime, we need norms to reduce the risks to which participants are exposed, and work has been done by researchers on this. Murray et al. ~\cite{murray2022ready} considered the legal problems that people need to face in a metaverse. Bokovnya et al.~\cite{bokovnya2020legal} discussed how realistic laws can be changed to combat cryptocurrency crime. Teichmann et al.~\cite{teichmann2020cryptocurrencies} later proposed a more effective international regulatory standard using the Liechtenstein Blockchain Act ~\cite{Liechtenstein} as a benchmark.

~\\

In addition to these three disciplines, there are many fields such as sociology, political science, international relations, etc. that are concerned with the changes that the metaverse may bring about, especially whether new financial crimes may evolve in such a ``beautiful new world'' as the metaverse. Since the day the financial markets were created, researchers and practitioners have been actively seeking strategies to combat financial crimes in various emerging areas in order to safeguard the smooth functioning of the system. Research into the financial aspects of the metaverse is still at an early stage and further exploration is needed in the future.

\subsection{Regulatory Policies and Measures}

As mentioned above, the new fertile soil of the metaverse has given birth to many new opportunities but is also coveted by many unscrupulous elements.
Many criminals have expanded their scams to the area of the metaverse. They take advantage of various loopholes in the still incomplete emerging technology to carry out attacks, causing many participating investors to lose their property.
Such financial crimes have largely undermined investors' confidence in the future of the metaverse, which is obviously not conducive to its long-term development. Therefore, government organizations around the world have started to introduce policies to regulate various digital assets and related services. 
\subsubsection{United States}
In the Anti-Money Laundering Act of 2020~\cite{AMLA} introduced in the United States, virtual assets and digital asset service providers have been included in the regulation of the Bank Secrecy Act. In March 2022, President Biden of the US signed a presidential order~\cite{bd}to ensure the responsible development of digital assets, which encourages regulators to monitor the risks posed by digital assets and develop policies to address vulnerabilities, support technological advances to ensure the security of digital assets
\subsubsection{Canada}
Canada has proposed stricter regulations for virtual currency trading. Cryptocurrency issuance service providers are considered as securities issuers and virtual currency dealers must register as money services businesses.
\subsubsection{European Union}
Cryptocurrencies in the EU are regulated by the Fifth Money Laundering Directive (5AMLD)~\cite{5AML} introduced back in 2020.It refers to the classification of cryptocurrencies and cryptocurrency exchanges as obligated entities,which involves customer service due diligence and suspicious activity reporting obligations. In addition, it gives financial intelligence units the power to obtain the address identities of owners of virtual currencies. The 5AMLD also proposes that exchanges and wallet providers need to be registered with the relevant domestic authorities, although this is in some sense contrary to the anonymity of cryptocurrencies.
\subsubsection{United Kingdom}
The UK government supports the circulation of cryptocurrencies. They endorse StableCoin as a means of payment, meanwhile, they have also proposed an economic crime legislative review process for crypto-assets. In November 2022, the UK financial regulator's joint statement~\cite{joint} on sanctions and the crypto-asset industry called on crypto-asset firms to identify customers and monitor transactions, update risk assessments of customers and transactions, and conduct reports in a timely manner. 
\subsubsection{Australia}
In Australia, crypto assets are regulated as financial products supervised by the Australian Securities and Investments Commission or as consumer products supervised by the Australian Competition and Consumer Commission. Crypto asset exchanges or crypto asset secondary service providers are required to register and be subject to AML/CFT regulation. 
\subsubsection{Singapore}
In Singapore, crypto assets are considered ``digital payment tokens", and crypto asset service providers are considered ``digital payment token services" and are both governed by the Payment Services Act.
Singapore recently passed the Financial Services and Markets Act 2022~\cite{singa}, which brings crypto asset companies located in Singapore but providing services outside of Singapore under the regulatory umbrella. Additionally, the bill introduces significant new licensing requirements and grants the Monetary Authority of Singapore new powers. 
\subsubsection{Japan}
Japan was one of the first countries in the world to introduce regulation of cryptocurrencies. Different types of tokens are governed by different regulatory provisions in Japan. Their regulatory status is documented in the Japanese Payment Services Law ~\cite{jap}. 

\subsubsection{China}
In China, the only cryptocurrency currently recognized as legal tender is the E-CNY, while trading in other cryptocurrencies is prohibited since September 2021. 
Overall, the regulation of digital assets is currently at an immature and experimental stage, with countries having different tools and measures.

\subsubsection{ICO Regulations}
In addition, Initial Coin Offering (ICO) scams have caused a lot of property losses, and countries have proposed related regulatory strategies for ICOs. The strictest countries should be China and South Korea, both of which have explicitly banned ICO activities.
In Russia, Thailand, and the Philippines, ICOs are not banned, but they are also regulated in a strict manner. Russia has set a cap on the number of funds raised for ICO projects (not to exceed 1 billion rubles), the Philippines' ICO activities need to be approved by the Philippine Securities and Exchange Commission. Similarly, ICO projects in Thailand are regulated by the authorities Thailand Securities and Exchange Commission , and ICO companies are obliged to provide the names of buyers and sellers and transaction information to the Anti-Money Laundering Office. In addition, the Thai tax authority will charge 7\% VAT and 15\% capital gains tax on cryptocurrencies and ICOs. Take Singapore as an example, there is also a part of the country that has always maintained a positive attitude towards ICOs. The Monetary Authority of Singapore only regulates related activities if the ICO poses specific risks.

As for the NFT and Defi programs, there are very few regulatory strategies implemented.
In summary, the regulation of crypto assets is still in a preliminary and immature stage, and the relevant agencies will intervene to improve security while inevitably reducing the anonymity and decentralization of transactions and curbing the free development of the industry to a certain extent. Therefore, how to reconcile the contradiction between the two is still a proposition that all countries in the world need to think about.

\section{Opportunities and Challenges}
\label{sec:OpportunitiesandChallenges}


In the previous section, we overview the research and measures related to coping with and preventing financial crimes in the metaverse economic system. At present, financial regulation and illegal behavior identification technologies related to the metaverse are mainly focused on the cryptocurrency level, and the research on regulation for the Web3 metaverse is still in its initial stage. However, we can foresee that, in the near future, the metaverse technology will play a fundamental role in a broader field, and the financial regulation on the Web3 metaverse will also show diversity and uniqueness of technology. Therefore, exploring the possible regulatory opportunities and challenges of the Web3 metaverse helps us propose ideas and insights to ensure the healthy development of the metaverse.
Since the Web3 metaverse integrates various emerging technologies and systems built on it as its foundation, the regulatory opportunities and challenges for them may also be inherited by the metaverse.
As mentioned earlier, the underlying technical foundation of the Web3 metaverse is blockchain technology, and thus the data of the Web3 metaverse also has the good nature of blockchain data: open and transparent, forgery-proof, tamper-proof, and traceable, which provides unprecedented opportunities for researchers to understand and solve related problems by analyzing blockchain data. Therefore, this section focuses on several opportunities and challenges of financial regulation in the metaverse from a data-driven perspective.

\begin{figure*}
    \centering
    \includegraphics[width=0.7\linewidth]{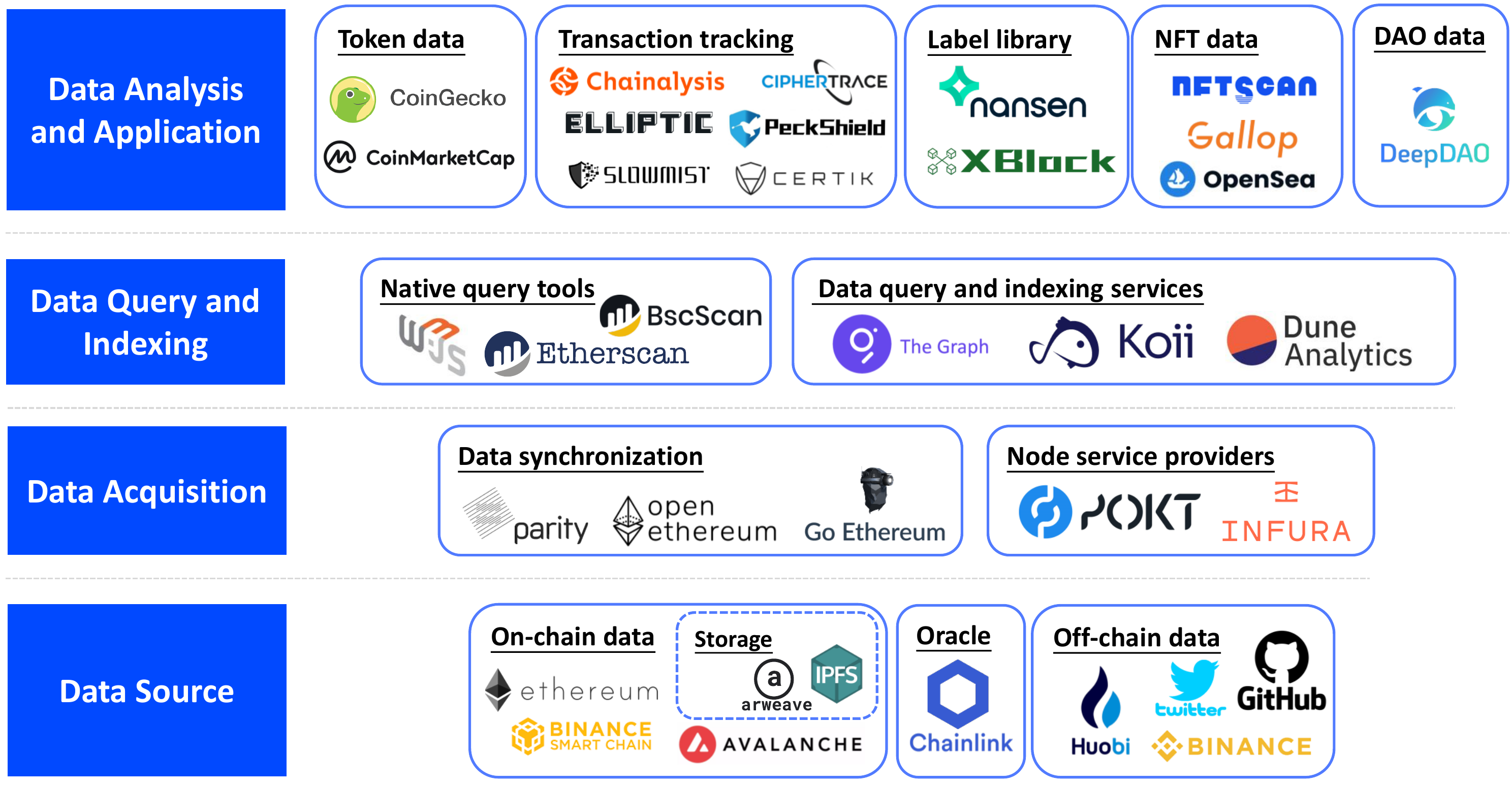}
    \caption{Web3 metaverse data track structure.}
    \label{fig:opportunitiesdata}
\end{figure*}

\subsection{Opportunities}
Next, we discuss the opportunities of data-driven financial governance in Web3 metaverse at four levels: the bottom level is data sources, the second level is data acquisition, the third level is data query and indexing, and the top level is data analysis and application, as shown in Figure~\ref{fig:opportunitiesdata}.

\subsubsection{Data Source}
\begin{figure}
    \centering
    \includegraphics[width=\linewidth]{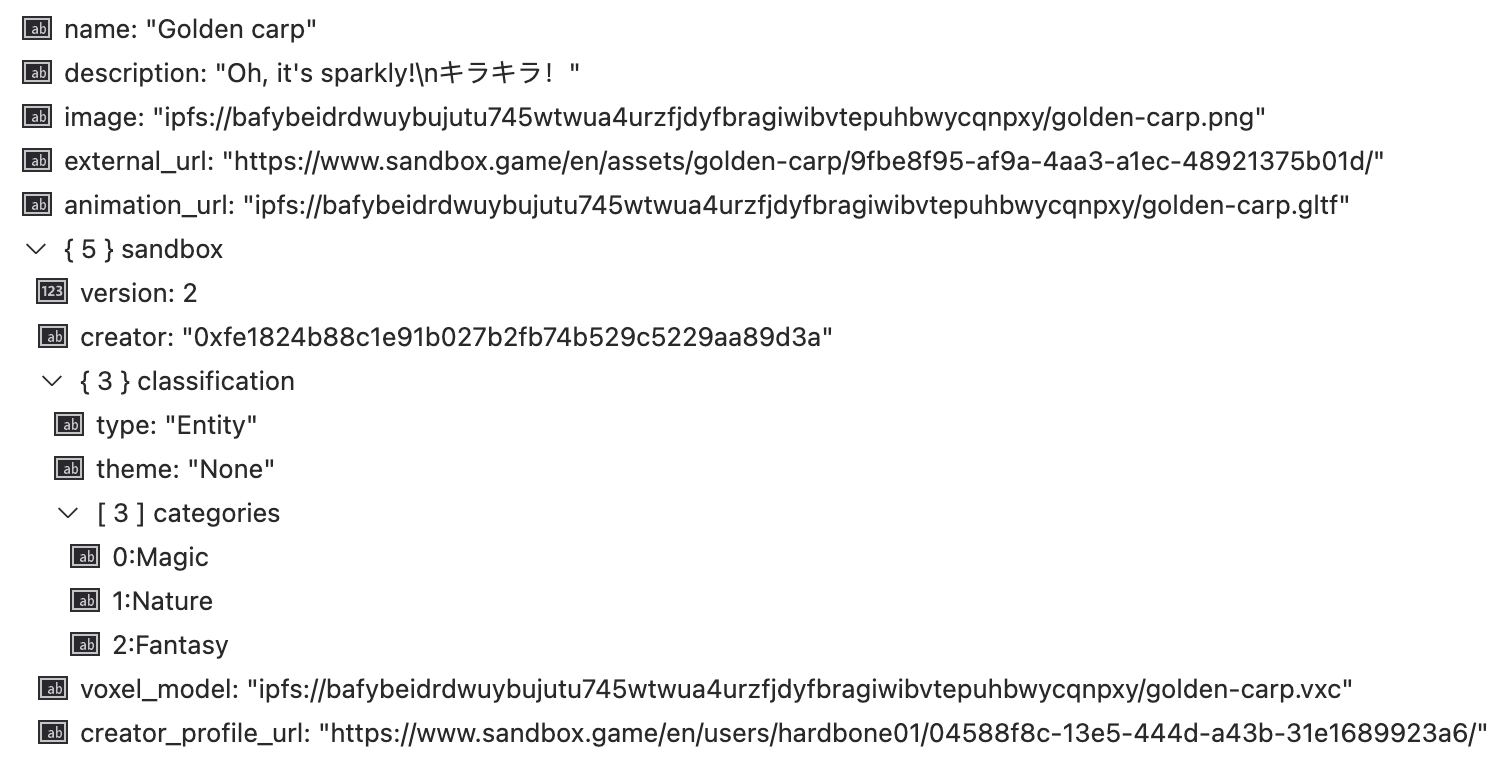}
    \caption{JSON file on IPFS of a digital asset named ``Golden carp'' on The Sandbox, a metaverse project.}
    \label{fig:IPFS_example}
\end{figure}

\begin{figure}
    \centering
    \includegraphics[width=\linewidth]{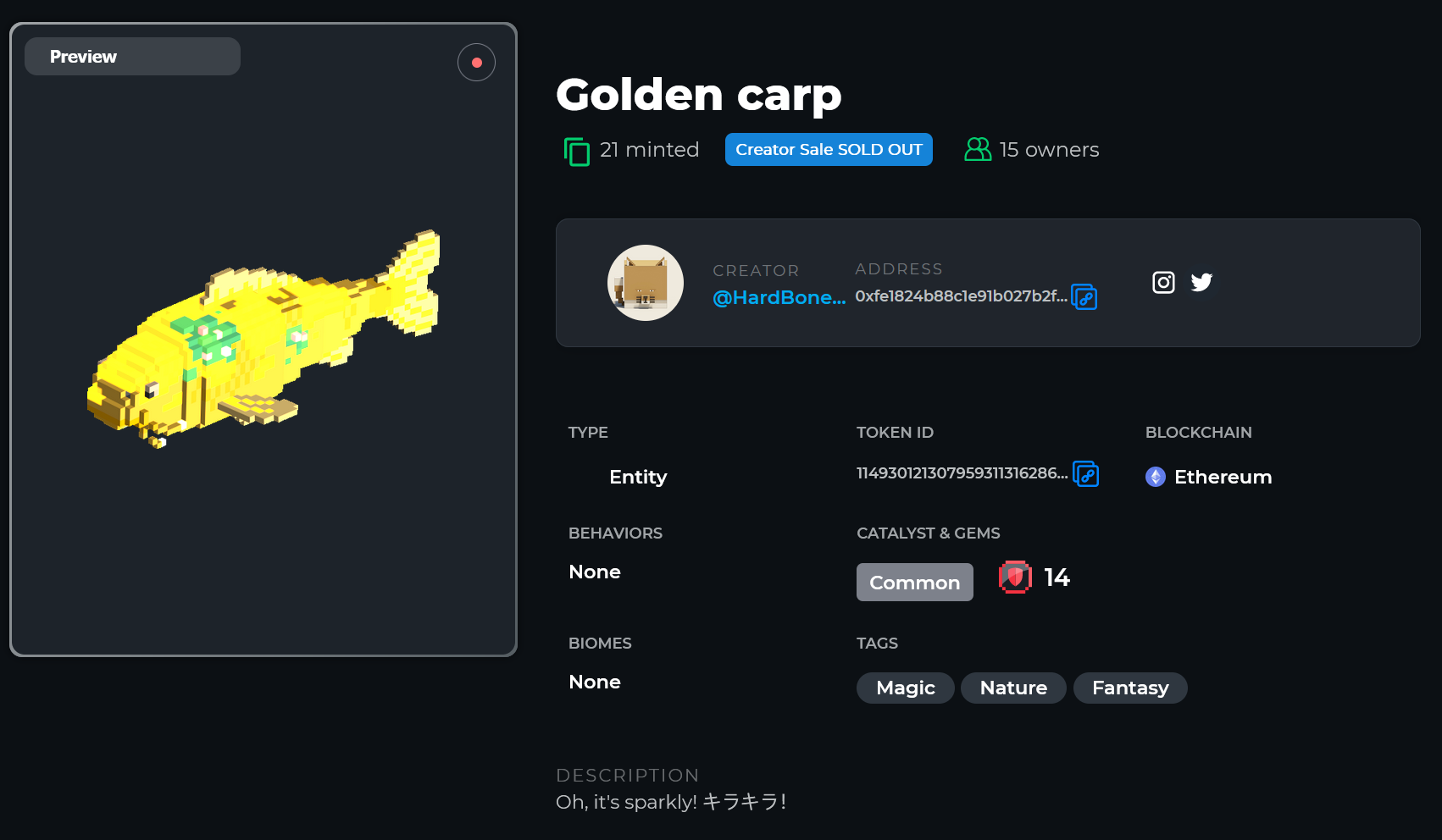}
    \caption{A digital asset on the Sandbox.}
    \label{fig:sandbox_example}
\end{figure}

The data of blockchain and Web3 metaverse can be categorized into on-chain data and off-chain data. On-chain data mainly includes blocks, transfer transactions, wallet addresses, smart contract bytecodes, smart contract events, digital asset information, and other data. In addition, decentralized storage is also the main source of on-chain data; for example, NFT can be saved in the Inter Planetary File System (IPFS)~\cite {benet2014ipfs}, a peer-to-peer hypermedia transfer protocol. The off-chain data, on the other hand, mainly includes data from centralized exchanges (e.g., Cryptocurrency Exchange), as well as some typical Web2 data, such as social media data, GitHub website data, etc.
For on-chain data we consider the metaverse project  Sandbox (\url{https://www.sandbox.game}) as an example. The Sandbox is a virtual metaverse built on the Ethereum blockchain where players can create, own, and monetize their gaming experiences. In the Sandbox game, the ERC-20 protocol is used to execute transactions, including tokens for custom avatars, land purchases, and general game interactions; ERC-1155 and ERC-721 are used to store and trade digital assets (including LAND, player-created ASSET). The Sandbox's NFT or DT transaction data is stored on the Ethereum public chain, and the rest of the NFT attribute information, such as images, is stored on IPFS, which is publicly accessible. Figure ~\ref{fig:sandbox_example} shows a digital asset on the Sandbox\footnote{\url{https://www.sandbox.game/en/assets/golden-carp/9fbe8f95-af9a-4aa3- a1ec-48921375b01d/}}, the asset's transaction and wallet address 0xa342f5d851e866e18ff98f351f2c6637f4478db5 associated with it. The rest of the information, including image, creator address, creator profile url, voxel model, etc., can be obtained from IPFS on json~\footnote{\url{https://ipfs.io/ipfs/ bafybeicliqwmhkue4d5ddrvdwhxpu3r3skh4kyxephwutz7dcywwle74ge/1.json}}, as shown in Figure ~\ref{fig:IPFS_example}.
For the off-chain data, we take the metaverse project Decentraland (\url{https://decentraland.org}) as an example.
Decentraland prepares resource information about technical components for community members or technicians, and it is open source on the GitHub site \url{https://github.com/decentraland/}. Moreover, the latest activities of the project can be found on its Twitter site (\url{ https://twitter.com/decentraland}).

\subsubsection{Data Acquisition}
There are more diverse ways to obtain Web3 metaverse data, but since off-chain data is often the data of centralized institutions (e.g. centralized trading platforms) or Web2 type data, the ways to obtain it vary greatly, so this paper mainly discuss the acquisition of on-chain data.

The underlying blockchain data of metaverse contains a large amount of heterogeneous data, and there are various methods to obtain Web3 metaverse on-chain data. Taking the Ethereum-based metaverse project as an example, the main ways to obtain Ethereum on-chain data include:
1) downloading and directly parsing block files. This method is simple and fast to implement, but it cannot collect complete data. This is because internal transactions are not stored in the blockchain and therefore cannot be obtained by parsing blocks; 2) deploying an Ethereum client, such as the Parity client's API allowing users to directly access internal transactions and external transactions in Ethereum. However, this method has limited access to data; for example, contracts' bytecode data and token transfer data are not accessible \cite{chen2019DataEther}.
In addition, these two methods require high time, money, and technical costs for users, and thus node service providers facilitating blockchain data acquisition have emerged. Node service providers can therefore also considered as the infrastructure for data analysis and mining in the Web3 metaverse. Some of the more well-known decentralized data service providers are Pocket Network (\url{https://www.pokt.network/}), whose core business is to provide decentralized data relay services (or RPC services) for developers on various public chains.
Pocket Network's decentralized infrastructure is helping Klaytn (an open-source meta-universe public chain project \footnote{\url{https://developer.klaytn.foundation/technology-overview/}}) realize its metaverse vision by driving Klaytn to new levels of decentralization, scalability and cross-chain capabilities.


\subsubsection{Data Query and Indexing}
\begin{figure*}
    \centering
    \includegraphics[width=\linewidth]{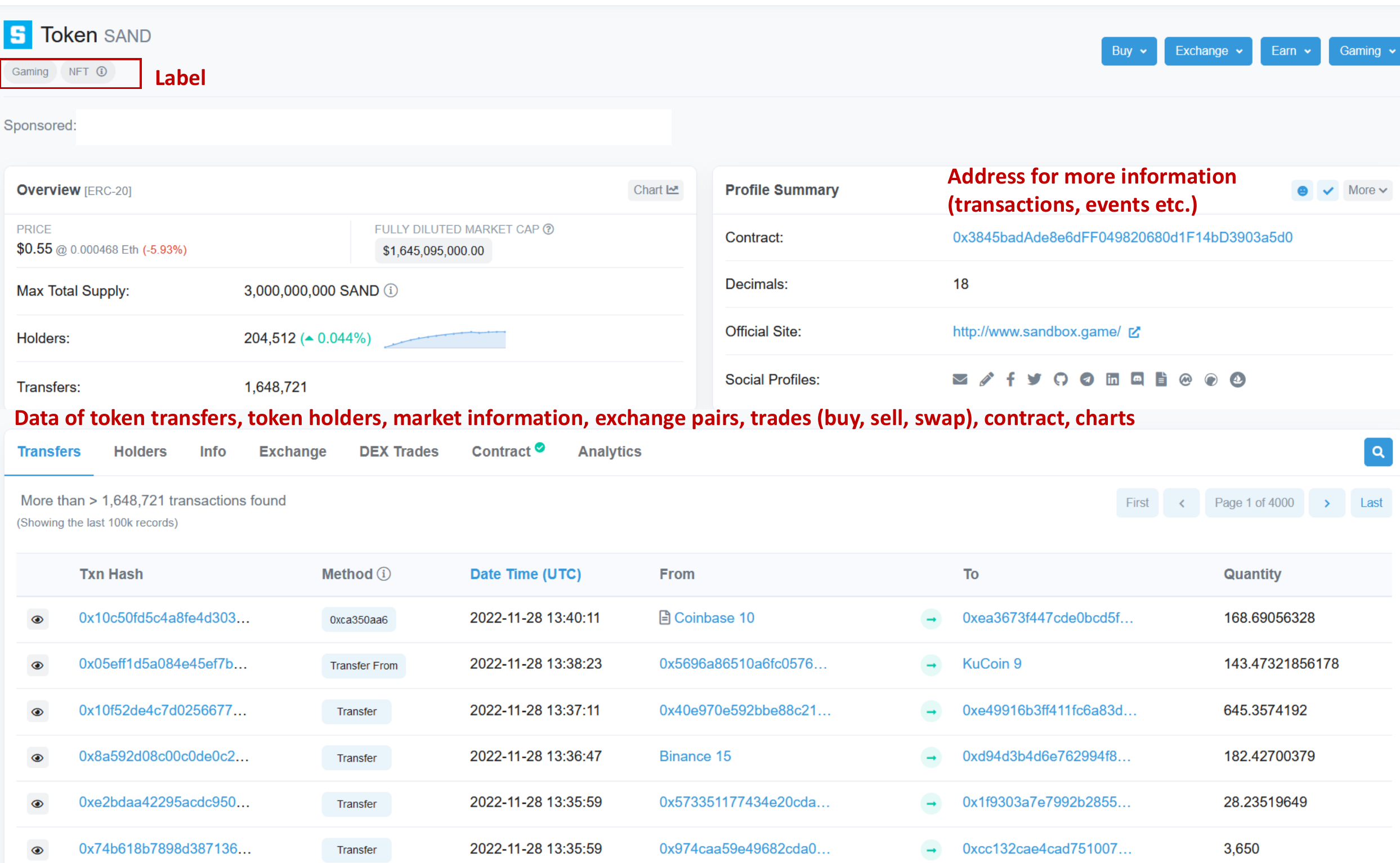}
    \caption{Metaverse project The Sandbox's SAND token on the Token Tracker page of Etherscan, the ethereum browser}
    \label{fig:SAND_token_tracker}
\end{figure*}

As mentioned earlier, almost all blockchain transactions in the metaverse are publicly available on the blockchain. However, it is usually a daunting work to query and index the raw transaction data, which is often large and diverse in data type.
The earliest tools for querying and indexing Web3 metaverse data were the APIs of the underlying public chain and blockchain browsers, such as the Web3 API and the Ethereum browser provided by Etherscan.
Web3 API uses a remote procedure call (RPC) to communicate with Etherscan nodes. For example, we call web3.eth.getTransaction() to get data of a specified hash transaction.
Etherscan, the Ethereum browser, allows users to search for information on the chain directly through the web page, including data on the chain, data on blocks, transaction data, smart contract data, address data, etc. As shown in Figure \ref{fig:SAND_token_tracker}, according to the address of Sandbox's SAND (traded and governed ERC20 tokens) 0 x3845badAde8e6dFF049820680d1F14bD3903a5d0, and its Token tracker~\footnote{\url{https://cn.etherscan.com/token/0 x3845badAde8e6dFF049820680d1F14bD3903a5d0}}, we can see a wealth of valuable information. This kind of data provides a database for web3 metaverse data mining and abnormal behavior identification.

In addition to the blockchain browsers of the underlying public chains of the metaverse, there are also service providers that offer data query and indexing services. They make raw data more accessible and usable by parsing and structuring it on top of node service providers that interact directly with various public chains.
In what follows, we give two representative examples: (1)
Dune Analytics (\url{https://dune.com/}) is a comprehensive Web3 data platform that adds raw data to SQL tables and parses them based on APIs provided by node service providers to enable users to query, analyze, and visualize through dashboards in real-time in their well-built databases using SQL. To date, over 10,000 analysts have created approximately 100,000 queries on Dune's platform, ranging from metrics for OpenSea NFTs to custom balance sheets for DAOs \footnote{\url{https://blockworks.co/news/web3-data-firm- dune-analytics-hits-unicorn-status}}.
A series of user-created metaverse-related dashboards can be seen on Dune Analysis, with one of the most popular items being Metaverse \& Virtual Real Estate\footnote{\url{https://dune.com/metaland/Metaverse-Land-Community}}, showing information such as Today's hottest metaverse land deals, metaverse rankings, average prices, and other key information. There is also a user-created NFT Wash Trader dashboard on Dune \footnote{https://dune.com/cryptok/NFT-Wash-Trading} dedicated to exposing the activity of LooksRare Wash Traders, including wallets, trades, and nft used, which provides NFT swipe detection technology research~\cite{von2022nft,Serneels2022,das2021understanding,tariq4097642suspicious} provides the technical foundation.
(2) The Graph (\url{https://thegraph.com}) is a decentralized on-chain data indexing protocol for querying networks like Ether and IPFS. The Graph is a cloud service like API composed of decentralized indexing nodes. The Graph can be accessed directly through subgraphs to obtain information, quickly and resource-efficiently. For example, Xia \textit{et al.}~\cite{Xia2021Trade} uses The Graph protocol to obtain trading events  associated with the decentralized exchange Uniswap (e.g., mint, swap, and burn events). Based on the data queried and parsed by The Graph, they deeply analyze the fraudulent tokens on Uniswap V2 and propose an effective and accurate method to detect fraudulent tokens, identifying over 10K fraudulent tokens and fraudulent liquidity pools, revealing the surprising fact that Uniswap is flooded with fraudulent tokens. This side-by-side demonstrates the necessity and feasibility of regulation in the decentralized financial ecosystem of the Web3 meta-universe.

\subsubsection{Data Analysis and Applications}

One layer up from the data query and indexing is the encapsulated, deliverable data products that can provide metaverse data value directly to users. The players in this layer can be broadly classified according to the type of data as token data analysis, on-chain transaction tracking, label library applications, NFT data analysis, DAO data analysis, etc.

\begin{itemize}
    \item \textbf{Token data analysis.} One representative platform is CoinMarketCap (\url{https://coinmarketcap.com}, established in 2013, which is used to observe and track token prices, trading volume, market value, etc. For example, CoinMarketCap gives a marketcap ranking of metaverse tokens (including trading tokens or governance tokens)\footnote{\url{https://coinmarketcap.com/view/metaverse/}} and a ``Play-to-Earn'' ranking of metaverse projects\footnote{\url{https://coinmarketcap.com/watchlist/6163287dad9db6359e33775b/}} for players' reference. In the area of regulation, Chen \textit{et al.}~\cite{chen2022cryptocurrency} conducted an empirical study on the analysis of cryptocurrency exchange swipes based on scores and rankings of exchanges provides by Coinmarketcap. Another data platform similar to CoinMarketCap is CoinGecko (\url{https://www.coingecko.com/}), which has been used by a number of researchers to conduct research on cryptocurrency and DeFi applications. For example, DeFiRanger~\cite{Wu2021DeFiRanger} refers to the market capacity and the price of tokens of five vulnerable DeFi apps provided by CoinGecko, to represent the market value of these DeFi apps, and proposes a price manipulation identification technique for DeFi apps. The Rug pull research~\cite{Mazorra2022} also makes reference to the CoinGecko, CoinMarketCap, and Etherscan data analysis websites and unearths 674 token lists marked as non-malicious to study and identify rug pull in DeFi applications. DeFiLlama (\url{https://defillama.com/}) is a popular aggregator of DeFi statistics cited in academic papers~\cite{meister2022yields} and official reports~\cite{eu2022defi} on DeFi. DeFiLlama also provides a list of hack incidents in the blockchain, DeFi, and cross-chain bridge space over the years\footnote{\url{https://defillama.com/hacks}} , including amount of loss, hacked public chains, attack categories, and other information. These classifications and statistics help researchers summarize and generalize the models of different types of attacks and design more accurate and efficient methods for DeFi attack detection and DeFi attack defense~\cite{Wang2021,Qin2021}.

    \item \textbf{On-chain transaction tracking.} The on-chain transaction tracking platform is a platform that has been around since the birth of Bitcoin. The representative platform is Chainalysis (\url{https://www.chainalysis.com/}), established in 2014 to help governments, cryptocurrency exchanges, international law enforcement agencies, banks, and other customers comply with compliance requirements, assess risk, and identify illegal activity through on-chain data monitoring and analysis. Research by companies such as Chainalysis has been an inspiration for the supervision of the entire ecosystem in metaverse. The Chainalysis Crime report~\cite{chainalysis2019}, which summarizes the crimes committed in 2018, provides a more in-depth look at the economic losses caused by phishing scams and provides case examples to inspire a series of subsequent papers on phishing scam detection~\cite{trans2vec2020Wu,Chen2020Phishing,yuan2020phishing,chen2021tegdetector,Li2021TTAGN,wen2022hide,xia2022phishing}. Recently, Chainalysis has revealed examples of NFT wash trading in its Web3 report~\cite{Chainalysis2022Web3}, which also provide inspiration for subsequent papers on NFT wash trading detection. Other on-chain transaction tracking platforms like Chainalysis include CiperTrace (\url{https://ciphertrace.com}), Elliptic (\url{https://www.elliptic.co}), PeckShield (\url{https://peckshield.com/}), and SlowMist. com/), SlowMist (\url{https://www.slowmist.com}), Certik (\url{https://www.certik.com}), etc. For blockchain forensics and on-chain transaction tracking tools, Srivasthav \textit{et al.}~\cite{Srivasthav2021Forensics} conduct a review of platforms and propose a taxonomy to map the identified advanced forensic features to the investigated supporting forensic tools. Transactions are the smallest unit of economic activity in the web3 metaverse, thus, on-chain transaction tracking platforms can effectively support the forensic and analytical work of financial crimes.

    \item \textbf{Label library applications.} The representative platform in industry is Nansen (\url{https://www.nansen.ai/}), founded in 2020. Nansen provides a number of wallet labels \footnote{\url{https://www.nansen.ai/guides/wallet-labels-emojis-what-do-they-mean}}, a way to tag and identify wallet addresses, classify wallets as ``Fund", ``Heavy DEX trader", ``Legendary NFT collector" etc. Mapping on-chain data with a database of millions of labels, researchers can understand what is happening on the blockchain in the metaverse and the types of wallets executing transactions, and can see who is behind these transactions. The representative platform in the academic community is Xblock (\url{http://xblock.pro/}), which provides several datasets that allow for transaction data analysis and anomalous behavior detection~\cite{Zheng2020XBlockETH,wu2022transaction,Lin2021Evolution,modeling2020lin,Wu2021JNCA}, including token price datasets, phishing scams datasets, and ponzi scams datasets. In addition to the datasets, the Xblock platform has launched XLabelCloud, an open label database that provides an online Chrome web plugin to facilitate researchers' investigations in scenarios such as blockchain browsers~\cite{chen2018detecting,jin2022time}. 

    \item \textbf{NFT data analysis.} Founded in 2021, the NFTscan platform (\url{https://www.nftscan.com/}) provides NFT collectors and investors and researchers with an API\footnote{\url{https://docs.nftscan.com/nftscan/API Overview}} to access NFT asset data and historical data held at any wallet address, as well as data analysis including top mint, gas tracker, NFT marketplace, trending collection, etc. NFTscan is designed to help users better track and evaluate the value of NFT assets to help make informed investment decisions. Such NFT data platforms have been used by researchers in academic studies, e.g., Cho \textit{et al.}~\cite{cho2022non} utilized data from six profile picture collections (PFP) type NFT collections provided by Gallop (\url{https://www.higallop.com/}), including transaction history, price, the associated wallet address, visual features and attachment of the NFT. The article also describes some of the challenges associated with NFT transaction data and data pre-processing recommendations. The dataset is currently open source~\cite{NFTdata2022ecommons}.

    \item \textbf{DAO data analysis.} As we know, in the process of on-chain decision making of a DAO, members first vote on the proposal on chain to decide whether to execute the proposal, and then the smart contract will automatically execute the proposal after the vote is passed. As the first DAO comprehensive data platform, DeepDAO (\url{https://deepdao.io/}), founded in 2020, analyzes, explores and ranks DAO based on multiple dimensions. For example, DeepDao provides an overview of the DAO of Decentraland\footnote{\url{https://dao.decentraland.org/en/}}, including information on project members' shares, proposals, Voting Coalitions, etc\footnote{\url{https://deepdao.io/organization/60c9b31c-4495-4028-aeac-eb7bb117fece/organization_data/members}}. Based on the data analysis of DeepDao, future researchers can explore the possible financial crimes in the metaverse DAO ecosystem, such as vote manipulation of DAOs, money laundering through DAOs, collusion or cronyism, and vote swiping of proposals.
    
\end{itemize}

As mentioned above, the underlying layer of the Web3 metaverse economic system is the blockchain. Due to the openness, traceability, and immutability of the blockchain, the transaction data, contract data, DAO members, and other data of the metaverse that contains rich information can be accessed publicly and completely. Meanwhile, based on various levels of Web3 metaverse data, many platforms are exploring and developing tools for data source, acquisition, query, indexing, and analysis, which provides unprecedented opportunities for data-driven research on metaverse financial regulatory technologies. The value of analyzing and mining Web3 metaverse financial data is twofold. 1) Researchers can broadly explore the evolution of user behavior, transaction networks, wealth distribution, asset values, and organizational decisions in the metaverse economic system, as a reference for other financial activities. 2) In recent years, various types of financial crimes have started to appear in the metaverse. Metaverse financial data analysis can help identify illegal behaviors among them and provide effective regulatory solutions for building a healthier metaverse ecology, and the related technology can also be a reference for metaverse transaction regulation in political affairs and other scenarios. In conclusion, these publicly available, processed, and easy-to-use data sources, access, query, and analysis platforms can not only enhance the theoretical value and application of data mining, social network analysis, quantitative trading, and other techniques in the financial system, but also help enhance the financial security and regulation of the meta-universe economic system.

\subsection{Challenges and Open Issues}

Although the publicly available data of the Web3 metaverse provides opportunities for technical research to prevent financial crimes, the ``decentralized" nature of Web3 also pose a great challenge for the governance of the metaverse. 
On the one hand, since the Web3 metaverse economic system integrates the latest technologies and systems such as blockchain, smart contracts, and digital assets as its foundation, the metaverse is very likely to inherit the regulatory challenges of these underlying technologies. 
On the other hand, the financial regulation of Web3 metaverse may face new challenges in the new scenario of metaverse.

\subsubsection{Challenges Introduced by Web3 Fundamentals}

From a technical point of view, Web3 provides the technical basis for the current hotly debated metaverse. From the economic point of view, compared with Web2, the most significant feature of Web3 is that it is a distributed Internet infrastructure, user-centered, emphasizing the autonomy of users' digital identity, personal data, and algorithms, and equal rights for users and builders.
According to the technical basis of Web3, the challenges of metaverse regulation brought by Web3 have three main levels:

\begin{itemize}
    \item \textbf{The underlying blockchain.} Different from traditional finance scenarios, blockchain is decentralized, borderless, and anonymous, while not limiting the number of accounts each user can open. Similar to blockchain, users of Web3 metaverse can conduct a large number of frequent transactions between accounts under their control. These result in the difficulty of identifying the entities of the Web3 metaverse accounts, with a large number of anonymous transactions and uncertain behavior. While de-anonymization may be possible through transaction data mining techniques, this in turn may raise other issues, such as user privacy breaches. Countless records of user activities and traces of user interactions will be retained in the Web3 metaverse. As these data are stored on the public blockchain, the accumulation of records and traces over time may cause user privacy disclosure problem.

    \item \textbf{Smart contracts and digital assets.} Smart contracts enable all types of digital assets, including stablecoins, fungible tokens, NFTs, etc. Smart contracts enable various types of digital assets to be exchanged on a trading platform. At the same time, Turing-complete smart contracts can represent and execute more complex application logic and functionality, leading to more complex transaction patterns. Meanwhile, as mentioned in the previous section, there are already many regulatory policies and measures for blockchain cryptocurrencies (e.g., Bitcoin, Ether, etc.) at home and abroad. The future regulation of the Web3 metaverse at home and abroad also needs to dovetail with the norms and measures for cryptocurrencies.
    
    \item \textbf{DeFi and DAO.} The goal of DeFi is to create a decentralized, open-source, permissionless and transparent economic system that operates behind a DAO~\cite{Salami2021} that operates strictly through programmed code/protocols. Although DeFi offers great opportunities for Web3 and the metaverse, DeFi and the DAOs behind it still need to be adequately regulated in order to ensure the trustworthiness of DeFi in the metaverse. However, users of current DeFi protocols or DApps are not mandated to meet anti-money laundering (AML) and know-your-customer (KYC) requirements. As described by Salami~\cite{Salami2021}, if a Web3 metaverse project has achieved a high degree of decentralization, they need to be operated and managed entirely by the DAO of the programming code/protocols without any influence from a centralized authority such as software developers. Then, it will become very difficult to hold anyone accountable for crimes and errors in the operation of the DeFi protocol in the Web3 metaverse.
    
\end{itemize}

\subsubsection{Open Issues Introduced by the Metaverse}

The metaverse constructs a new social structure where the virtual and the real are highly intertwined. The users of the metaverse are also residents of the real world, thus also making traditional security risks and non-traditional security risks superimposed on each other, and the virtual economy of the metaverse and the real economy of the real world will inevitably interact with each other. In this part, we will discuss the open issues introduced by the metaverse to financial regulation in terms of the different paths into the metaverse.

\begin{itemize}
    \item \textbf{Digital twins.} A digital twin is a digital mapping of the physical world, where the user enters the metaverse with their digital body. At this point, the definition of personal identity becomes problematic. In the real world, financial regulation regulates the actual person. It is then a challenge to regulate the interaction between this digital avatar and the real world person in person in a metaverse regulatory regime.

    \item \textbf{Digital primordial.} The digital native is a virtual universe parallel to the physical world, where multiple avatars of the self in the metaverse can multitask, collaborate and talk to each other. Therefore, a criminal in the real world can have multiple doppelgangers in the metaverse, and it will be difficult to correspond between the metaverse and the real-world ``person" in fact. At the same time, the metaverse breaks through national geographical boundaries, which poses a major obstacle to effective financial regulation and enforcement in individual countries.

    \item \textbf{Virtual-real synthesis.} The real world interacts with the virtual world. A real-world criminal can take the assets he illegally obtained in the metaverse (e.g., stealing assets via DeFi exploits) and can exchange the stolen assets on the chain for real-world fiat currency through an exchange that does not require KYC. The laundered fiat currency, in turn, flows into the real-world financial system and may be used to finance real-world terrorism. Thus, the interconnection of the real world and virtual world in the metaverse makes the metaverse economic system will face more severe risks than the traditional financial industry, which puts higher demands on the risk-awareness of the metaverse financial system.
    
\end{itemize}

\subsection{A Vision for the Future Regulation}
The Web3 metaverse relies on decentralized autonomous organizations (DAOs). For the future governance and supervision of Web3 metaverse, it may be possible to use the DAOs of metaverse to encode the operation and governance rules of metaverse in the form of smart contracts on the blockchain, and establish reward and punishment measures with the help of tokens, NFTs and other economic rights to finally realize the autonomous governance and autonomous supervision of Web3 metaverse. 
However, regulation and governance rules of DAOs may not cover all the risks and security issues in the metaverse.
Therefore, the government should pay high attention to crimes against the metaverse, prevent them from happening, guide the development of a set of universally applicable industry standards for the metaverse, and implement blockchain-based industry autonomy. At the same time, the government should cultivate industry consensus and optimize the regulatory scheme in the governance process in order to avoid the various drawbacks caused by brutal developments followed by a strong governance.

\section{Conclusion}
\label{sec:Conclusion}



This paper focuses on financial crimes in Web3-empowered metaverse. First, we introduce the background, foundation and diverse applications of the Web3 
 metaverse economic system. Then, we summarize the financial crimes and anti-crime techniques that have emerged on the metaverse from both academia, industry and government. Particularly, this paper provides a taxonomy of financial crimes on the metaverse, and a specific discussion of each crime, including existing definitions of the crime, case studies and analyses related to the metaverse, and existing academic research on the crime. Finally, the paper explores the possible opportunities and challenges of data-driven metaverse regulation. Overall, by providing an overview of this paper, we hope that readers can gain a better understanding of the financial crime issues that the metaverse may currently face and that it will help to improve metaverse regulation.
Among the future research directions, researchers may be able to explore the following directions:
1) Researchers can synthesize and analyze the current case data as well as employ corresponding statistical and analytical methods to explore the prevalence and trends of financial crimes in the metaverse and the corresponding countermeasures. 2) Researchers can also utilize the latest technologies and tools, such as artificial intelligence and big data analysis, to support research efforts on financial crimes in the metaverse.
3) Experts and scholars from cross-disciplinary disciplines work together to conduct in-depth research and analysis of current domestic and international legal and regulatory rules, and propose policies and recommendations that are more suitable for the long-term development of the metaverse.

\newpage

\bibliographystyle{IEEEtran}  
\bibliography{ref-1109}

\begin{IEEEbiography}[{\includegraphics[width=1in,height=1.25in,clip,keepaspectratio]{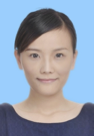}}]{Jiajing Wu}~(Senior Member, IEEE) received the Ph.D. degree from The Hong Kong Polytechnic University, Hong Kong, in 2014. In 2015, she joined Sun Yat-sen University, Guangzhou, China, where she is currently an Associate Professor. Her research interests include blockchain, graph mining, and network science. Dr. Wu was the recipient of the Hong Kong Ph.D. Fellowship Scheme during her Ph.D. in Hong Kong from 2010 to 2014. She is also an Associate Editor for {\sc IEEE Transactions on Circuits and Systems II: Express Briefs.}
\end{IEEEbiography}

\begin{IEEEbiography}[{\includegraphics[width=1in,height=1.25in,clip,keepaspectratio]{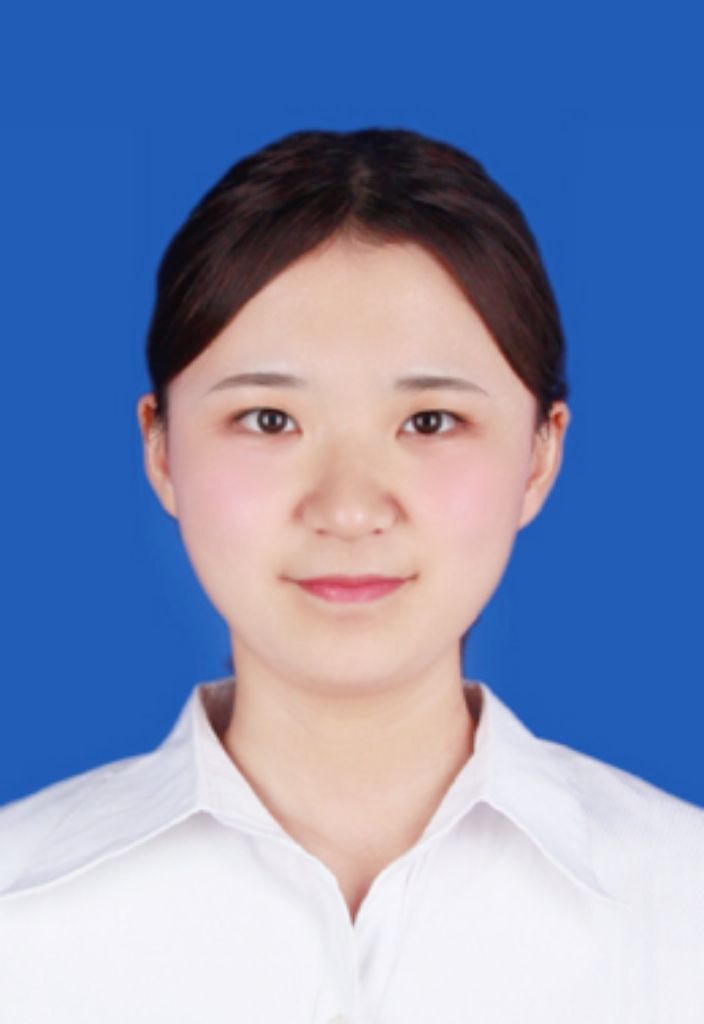}}]{Kaixin Lin}~received her B.Eng. in School of Computer Science from South China Normal University, Guangzhou, China, in 2022.
She is currently studying toward the M.Sc. degree in the School of Computer Science and Engineering, Sun Yat-sen University. Her current research interests include blockchain, smart contracts, and graph mining.
\end{IEEEbiography}

\begin{IEEEbiography}[{\includegraphics[width=1in,height=1.25in,clip,keepaspectratio]{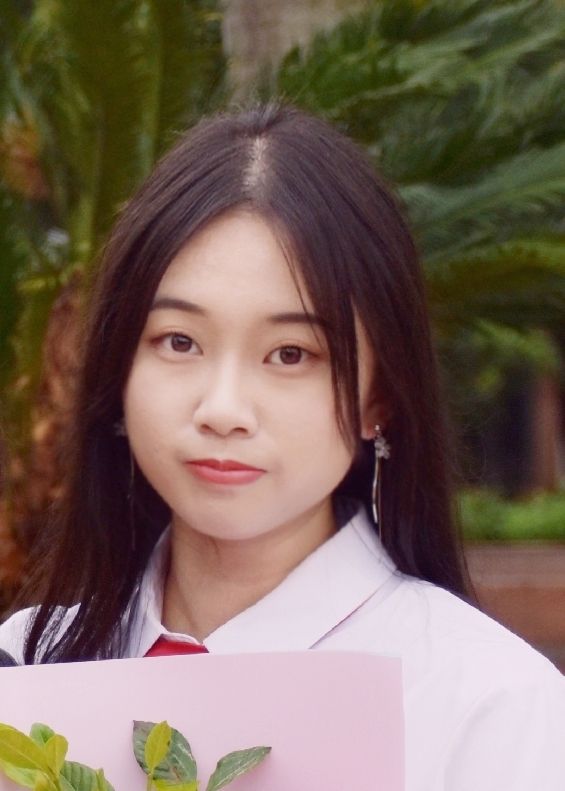}}]{Dan Lin}~(Graduate Student Member, IEEE) received her B.Eng. in Software Engineering from Sun Yatsun University, Guangzhou, China, in 2019. She is currently studying toward the Ph.D. degree in the School of Software Engineering, Sun Yat-sen University. Her current research interests include blockchain, cryptocurrency, theories and applications of network science, and anti-money laundering.
\end{IEEEbiography}

\begin{IEEEbiography}[{\includegraphics[width=1in,height=1.25in,clip,keepaspectratio]{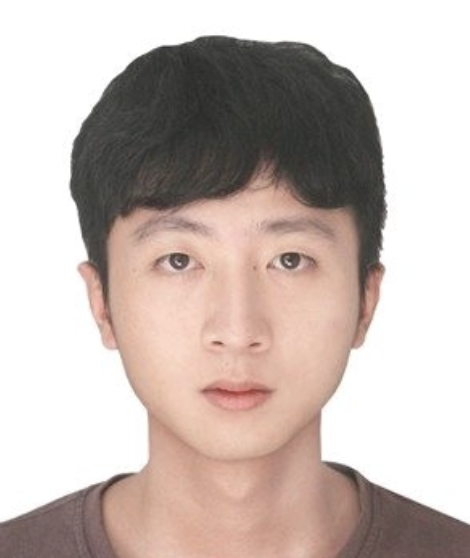}}]{Ziye Zheng}~is currently studying toward the B.Eng. degree in the School of Software Engineering, South China Normal University, Foshan, China. He is currently a research assistant in the School of Computer Science and Engineering, Sun Yat-sen University.
His research interests include blockchain, cryptocurrency, and graph mining.
\end{IEEEbiography}

\begin{IEEEbiography}[{\includegraphics[width=1in,height=1.25in,clip,keepaspectratio]{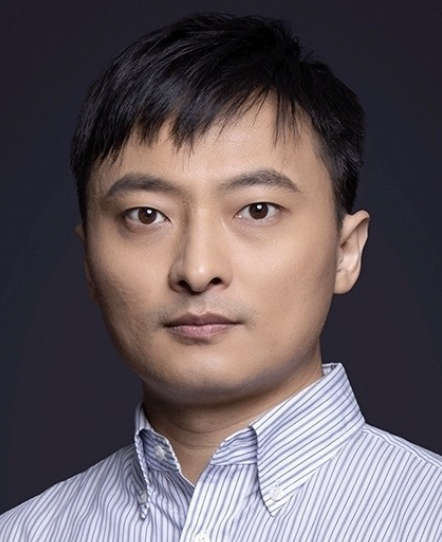}}]{Huawei Huang}~(Senior Member, IEEE) received the Ph.D. degree in computer science and engineering from the University of Aizu,
Aizuwakamatsu, Japan, in 2016. He is currently an Associate Professor with Sun Yat-Sen University, Guangzhou, China. He was a Research Fellow of Japan Society for the Promotion of Science, and an Assistant Professor with Kyoto University, Kyoto, Japan. His research interests include blockchain and distributed computing. He is also a Guest Editor of the {\sc IEEE Journal on Selected Areas in Communications and IEEE Open Journal of the Computer Society}, Operation-Committee Chair of the {\sc IEEE Symposium on Blockchain at IEEE Services 2021}, and the TPC Co-Chair of {\sc Globecom'2021/ICC'2022 Workshop on Scalable, Secure, and Intelligent blockchain}.
\end{IEEEbiography}
\begin{IEEEbiography}[{\includegraphics[width=1in,height=1.25in,clip,keepaspectratio]{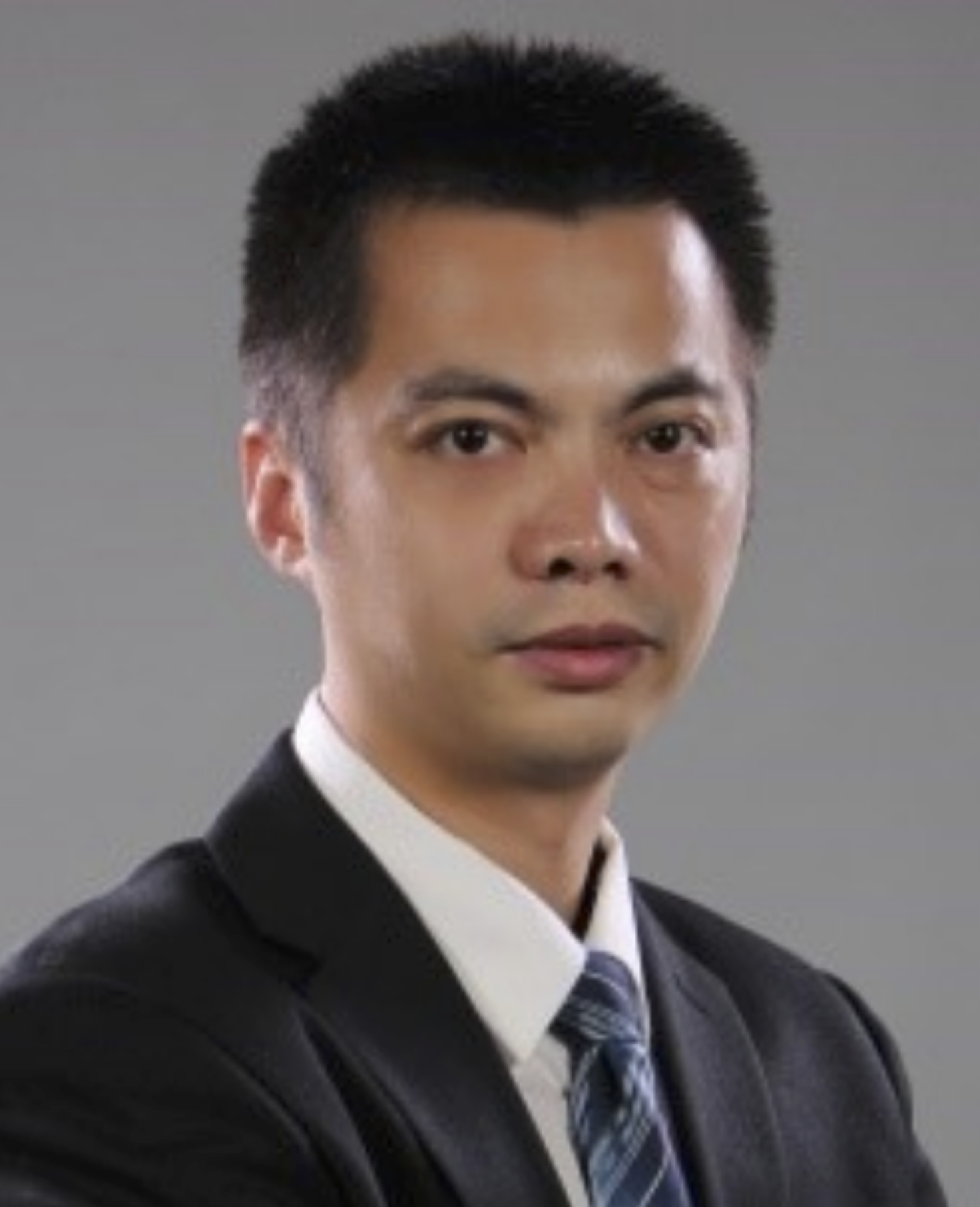}}]{Zibin Zheng}~(Fellow, IEEE) is currently a Professor and the Deputy Dean with the School of Software Engineering, Sun Yat-sen
University, Guangzhou, China. He authored or coauthored more than 200 international journal and conference papers, including one ESI hot
paper and six ESI highly cited papers. According to Google Scholar, his papers have more than 15 000 citations. His research interests include
blockchain, software engineering, and services computing. He was the {\sc BlockSys’19} and {\sc ccollaboratecom16} General Co-Chair, {\sc SC2’19}, {\sc ICIOT18} and {\sc IOV14} PC CoChair. He was the recipient of several awards, including the Top 50 Influential Papers in Blockchain of 2018, the {\sc ACM SIGSOFT} Distinguished Paper Award at {\sc icse2010}, the Best Student Paper Award at {\sc ICWS2010}.
\end{IEEEbiography}

\end{document}